\documentclass[twocolumn,english,preprintnumbers,amsmath,amssymb,aps,nofootinbib]{revtex4}
\usepackage[T1]{fontenc}
\usepackage[latin9]{inputenc}
\usepackage{color}
\usepackage{amsmath}
\usepackage{amssymb}
\usepackage{graphicx}
\usepackage{esint}
\usepackage{soul}

\makeatletter

\providecommand{\tabularnewline}{\\}
\usepackage{dcolumn}
\usepackage{bm}
\usepackage{amsfonts}
\usepackage{babel}
\providecommand{\U}[1]{\protect\rule{.1in}{.1in}}
\usepackage{ragged2e}
\usepackage{hyperref}

\newcommand{\del}{\partial}
\newcommand{\grad}{\nabla}

\newcommand{\scri}{\mathcal{I}}

\newcommand{\hateq}{\mathrel{\mathop {\widehat=} }} 

\makeatother

\begin{document}
	\title{Gravitational radiation with $\Lambda>0$}
	\author{B\'eatrice Bonga}
	\email{bbonga@science.ru.nl}
	\affiliation{Institute for Mathematics, Astrophysics and Particle Physics, Radboud
		University, 6525 AJ Nijmegen, The Netherlands}
	
	\author{Claudio Bunster}
	\email{bunster@cecs.cl}
	\affiliation{Centro de Estudios Cient\'ificos (CECs), Av. Arturo Prat 514, Valdivia,
		Chile\\}
	\affiliation{Universidad San Sebasti\'an, Chile}
	
	\author{Alfredo P\'erez}
	\email{alfredo.perez@uss.cl}
	\affiliation{Centro de Estudios Cient\'ificos (CECs), Av. Arturo Prat 514, Valdivia,
		Chile\\}
	\affiliation{Facultad de Ingenier\'ia, Arquitectura y Dise\~no, Universidad San Sebasti\'an, sede Valdivia, General Lagos 1163, Valdivia 5110693, Chile}
	
	\date{\today}
	\begin{abstract}
		We study gravitational radiation for a positive value of the cosmological constant $\Lambda$. We
		rely on two battle-tested procedures: (i) We start from the same null coordinate system used by Bondi and Sachs for $\Lambda = 0$, but, introduce boundary conditions adapted to allow radiation when $\Lambda>0$. (ii) We determine the asymptotic symmetries by studying, \`a la Regge-Teitelboim, the surface integrals generated in the action by these boundary conditions. 
		A crucial difference with the $\Lambda=0$ case is that the wave field does not vanish at large distances, but is of the same order as de Sitter space.
		This novel property causes no difficulty; on the contrary, it makes quantities finite at every step, without any regularization.
		A direct consequence is that the asymptotic symmetry algebra consists only of time translations and space rotations. Thus, it is not only finite-dimensional, but smaller than de Sitter algebra.
		We exhibit formulas for the energy and angular momentum and their fluxes. In the limit of $\Lambda$ tending to zero, these formulas go over continuously into those of Bondi, but the symmetry jumps to that of Bondi, Metzner and Sachs. The expressions are applied to exact solutions, with and without radiation present, and also to the linearized theory. 
	\end{abstract}
	\maketitle
	
	\section{Introduction}
	
	We study gravitational radiation for a positive value of the cosmological constant $\Lambda$ as generated by compact sources such as stars and black holes. We are guided by, and closely follow
	the steps of, Sachs' classical analysis of the concepts introduced
	by Bondi for $\Lambda=0$ \citep{Bondi:1960jsa,Sachs:1962wk,Sachs:1962zza}. That analysis gave unambiguous expressions
	for energy and energy flux, and also established the existence of
	an infinite-dimensional symmetry algebra, now called the Bondi-Metzner-Sachs (BMS) algebra.
	
	Bondi and Sachs relied only on Einstein's equations with  boundary conditions to study a large class of spacetimes, now known as ``asymptotically flat'' solutions. They did not, and needed not to, invoke the action principle. They required though, that the mass should decrease when radiation is emitted. This demand was essential
	to arrive at unambiguous expressions for the mass and its flux. We have not been able to implement the mass diminution requirement for $\Lambda>0$, although we do recover this in the linearized case. In order to arrive at the formulas for the mass and
	its flux, we have appealed instead to the action principle, including in it appropriate surface integrals \`a la Regge-Teitelboim.

	The symmetry algebra consists only of time translations and space rotations, even when gravitational radiation is present. It is not only finite-dimensional but smaller than the de Sitter algebra. 
	This result is crucially linked to the $\mathbb R \times \mathbb S^2$ topology of the future boundary, and 
	as shown in \citep{abk1}, the symmetry group for asymptotically de Sitter spacetimes depends crucially on the topology.
	The existence of this smaller symmetry algebra can be attributed to the fact that in the presence of a positive $\Lambda$ the wave field does not vanish at large distances, in sharp contrast with the asymptotically flat case. As a result, a generic gravitational wave will induce a strong deformation on the geometry of the future boundary. This deformation of the asymptotic region precludes the presence of the full de Sitter group or infinite-dimensional extensions thereof.
	In the limit $\Lambda \rightarrow 0$, the mass and flux formulas coincide with those of Bondi and Sachs. In contradistinction, the symmetry algebra has an enormous jump: It becomes the BMS symmetry. 
	
	For waves of small amplitude, we recover the energy
	flux of the linear theory on a de Sitter background. Interestingly, even waves with small amplitudes reach infinity without decay. This is not the case for asymptotically flat spacetimes, in which linear waves decay at least as $1/r$ at large distances. For special solutions such as Kerr-de Sitter and Robinson-Trautman with $\Lambda>0$, we recover the accepted expressions for the mass and angular momentum.
	All of our conclusions follow from General Relativity, by bringing into it boundary conditions which are the natural extension for $\Lambda>0$ of those employed by Sachs for $\Lambda=0$.
	
	Interestingly, all equations in this paper hold also for $\Lambda<0$. In that case, the boundary conditions do not correspond to the typical reflecting ones (see e.g.~\citep{Ashtekar:1984zz,Henneaux:1985tv,Ashtekar:1999jx}).\footnote{This is clear from the non-conformal flatness of the boundary metric.}  We find this an appealing issue for exploration, but we will not address it here.

	The boundary conditions in this paper are new and have not been explored elsewhere. These new boundary conditions have a finite-dimensional symmetry algebra, lead to finite charges and accommodate many interesting spacetimes describing radiation in de Sitter-like spacetimes (such as Robinson-Trautman with $\Lambda>0$ and linearized waves on a de Sitter background).
	Some of these properties have been obtained from different boundary conditions and/or different methods as well. We compare our results to other approaches in Sec.~\ref{sec:comparison}.
	
	The set-up of this paper is as follows. We introduce the Bondi-Sachs coordinates and their fall-off rates in Sec.~\ref{sec:BS-coordinates}. Given these coordinates, we study the asymptotic symmetry algebra in Sec.~\ref{sec:symmetries}. In Sec.~\ref{sec:charges-and-fluxes}, we compute the energy and angular momentum at future infinity and their fluxes. We apply this formalism to various exact solutions of General Relativity with $\Lambda>0$ in Sec.~\ref{sec:examples}, in which we also study linearized gravitational waves on a de Sitter background. Finally, we compare our approach to various alternative approaches in Sec.~\ref{sec:comparison}.
	The key findings of this paper are summarized in Table~\ref{tab:table1}.
	
	\begin{table*}
		\protect\caption{\label{tab:table1} Key results summarized. Cases $\Lambda=0$ and $\Lambda >0$ compared and contrasted.}
		\footnotesize{
			\begin{ruledtabular} %
				\begin{tabular}{llc|c}
					& \multicolumn{1}{c}{$\Lambda=0$} & \multicolumn{2}{c}{$\Lambda>0$}
					\tabularnewline
					Asymptotic region & \multicolumn{1}{c}{Future null $\scri$} & \multicolumn{2}{c}{Future space-like $\scri$}
					\tabularnewline
					\tabularnewline
					Topology asymptotic region & \multicolumn{1}{c}{$\mathbb{R} \times \mathbb{S}^2$} & \multicolumn{2}{c}{$\mathbb{R} \times \mathbb{S}^2$ (= $\mathbb{S}^3$ with two points removed)}
					\tabularnewline
					& \multicolumn{1}{c}{} & \multicolumn{2}{c}{to describe radiation emitted by bounded sources}
					\tabularnewline
					\tabularnewline
					Conformal completion & \multicolumn{1}{c}{Yes} & \multicolumn{2}{c}{Yes}\tabularnewline
					of infinity possible? &  & \multicolumn{2}{l}{}\tabularnewline
					&  & \multicolumn{2}{l}{}\tabularnewline
					Coordinates & \multicolumn{1}{l}{$ds^{2}=e^{2\beta}\frac{V}{r}du^{2}-2e^{2\beta}dudr$} & \multicolumn{2}{l}{$ds^{2}=e^{2\beta}\frac{V}{r}du^{2}-2e^{2\beta}dudr$}\tabularnewline
					& $\qquad+r^{2}g_{AB}\left(dx^{A}-U^{A}du\right)\left(dx^{B}-U^{B}du\right)$, & \multicolumn{2}{l}{$\qquad+r^{2}g_{AB}\left(dx^{A}-U^{A}du\right)\left(dx^{B}-U^{B}du\right)$,}\tabularnewline
					& $\det g_{AB}=\sin^{2}\theta$ & \multicolumn{2}{l}{$\det g_{AB}=\sin^{2}\theta$}\tabularnewline
					&  & \multicolumn{2}{l}{}\tabularnewline
					& (Bondi gauge with the Sachs condition) & \multicolumn{2}{l}{(Bondi gauge with the Sachs condition)}\tabularnewline
					&  & \multicolumn{2}{l}{}\tabularnewline
					&  & \multicolumn{2}{c}{}\tabularnewline
					Fall-off & $\beta=O\left(r^{-2}\right)$, & \multicolumn{2}{l}{$\beta=O\left(r^{-2}\right)$,}\tabularnewline
					&  & \multicolumn{2}{l}{}\tabularnewline
					& $U^{A}=-\frac{1}{2r^{2}}D_{B}C^{AB}$ & \multicolumn{2}{l}{$U^{A}=U_{\left(0\right)}^{A}-\frac{1}{2r^{2}}D_{B}C^{AB}$}\tabularnewline
					& $\qquad-\frac{2}{3r^{3}}\left(N^{A}-\frac{1}{2}C^{AB}D^{C}C_{BC}\right)+\dots$, & \multicolumn{2}{l}{$\qquad-\frac{2}{3r^{3}}\left(N^{A}-\frac{1}{2}C^{AB}D^{C}C_{BC}\right)+\dots$,}\tabularnewline
					&  & \multicolumn{2}{l}{}\tabularnewline
					& $V=-r+2M+\dots$, & \multicolumn{2}{l}{$V=\frac{\Lambda r^{3}}{3}-D_{A}U_{\left(0\right)}^{A}r^{2}$$-\left(1+\frac{\Lambda}{16}C_{AB}C^{AB}\right)r+2M+\dots,$}\tabularnewline
					&  & \multicolumn{2}{l}{}\tabularnewline
					& $g_{AB}=\gamma_{AB}+\frac{C_{AB}}{r}+\frac{\gamma_{AB}C_{CD}C^{CD}}{4r^{2}}+\frac{E_{AB}}{r^{3}}+\dots$ & \multicolumn{2}{l}{$g_{AB}=\gamma_{AB}+\frac{C_{AB}}{r}+\frac{\gamma_{AB}C_{CD}C^{CD}}{4r^{2}}+\frac{E_{AB}}{r^{3}}+\dots$,}\tabularnewline
					&  & \multicolumn{2}{l}{}\tabularnewline
					&  & \multicolumn{2}{l}{$D_{A}U_{\left(0\right)B}+D_{B}U_{\left(0\right)A}-\gamma_{AB}D_{C}U_{\left(0\right)}^{C}=\frac{\Lambda}{3}C_{AB}$}\tabularnewline
					&  & \multicolumn{2}{l}{}\tabularnewline
					Radiation field vanishes & \multicolumn{1}{c}{Yes } & \multicolumn{2}{c}{No }
					\tabularnewline
					at infinity & \multicolumn{1}{c}{ ($U^A_{(0)}=0$)} & \multicolumn{2}{c}{ ($U^A_{(0)} \neq 0$)}
					\tabularnewline
					\tabularnewline
					Imprint on the & Symmetric traceless tensor $C_{AB}$  & \multicolumn{2}{l}{Symmetric traceless tensor $C_{AB}$}\tabularnewline
					metric of the most & arbitrary functions of the retarded time and & \multicolumn{2}{l}{arbitrary functions of the retarded time and}\tabularnewline
					general wave & the angles (generic graviton) & \multicolumn{2}{l}{the angles (generic graviton)}\tabularnewline
					&  & \multicolumn{2}{l}{}\tabularnewline
					Symmetry & Infinite-dimensional (``BMS'') Lie algebra: & \multicolumn{2}{l}{Four-dimensional Lie algebra:}\tabularnewline
					& $so\left(3,1\right)+\text{``supertranslations''}$ & \multicolumn{2}{l}{ $so\left(3\right)\oplus\mathbb{R}$}\tabularnewline
					&  & \multicolumn{2}{l}{}\tabularnewline
					Energy (``Bondi mass'') & \multicolumn{1}{l}{$E=\frac{1}{4\pi G}\oint d^{2}S\,M$} & \multicolumn{2}{l}{$E=\frac{1}{4\pi G}\oint d^{2}S\,M$}\tabularnewline
					&  & \multicolumn{2}{c}{}\tabularnewline
					Angular momentum & \multicolumn{1}{l}{$\vec{J}=\frac{1}{8\pi G}\oint d^{2}S\,\hat{r}\epsilon^{AB} D_{A}N_{B}$} & \multicolumn{2}{l}{$\vec{J}=\frac{1}{8\pi G}\oint d^{2}S\,\hat{r} \epsilon^{AB} D_{A}N_{B}$}\tabularnewline
					&  & \multicolumn{2}{c}{}\tabularnewline
					Angular momentum & \multicolumn{1}{c}{Yes} & \multicolumn{2}{c}{No}\tabularnewline
					ambiguity & \multicolumn{1}{c}{(angular momentum not invariant} & \multicolumn{2}{c}{(there are no supertranslations)}\tabularnewline
					& \multicolumn{1}{c}{under supertranslations)} & \multicolumn{2}{c}{}\tabularnewline
					&  & \multicolumn{2}{c}{}\tabularnewline
					Energy flux & $\frac{dE}{du}=-\frac{1}{32\pi G}\oint d^{2}S\,N_{AB}N^{AB},$ & \multicolumn{2}{l}{$\frac{dE}{du}=-\frac{1}{32\pi G}\oint d^{2}S\,\left[N_{AB}^{\left(\Lambda\right)}N^{\left(\Lambda\right)AB}+\frac{2\Lambda}{3}C^{AB}C_{AB}\right.$}\tabularnewline
					& with $N_{AB}:=\dot{C}_{AB}$ & \multicolumn{2}{c}{$\qquad-\frac{\Lambda}{6}C^{AB} D^2 C_{AB}+\frac{7\Lambda^{2}}{144}\left(C^{AB}C_{AB}\right)^{2}-\frac{\Lambda^{2}}{3}C^{AB}E_{AB}$}\tabularnewline
					&  & \multicolumn{2}{l}{$\qquad\left.+\left(4M+D_{A}D_{B}C^{AB}\right)\left(D_{C}U_{\left(0\right)}^{C}\right)\right],$}\tabularnewline
					&  & \multicolumn{2}{c}{}\tabularnewline
					&  & \multicolumn{2}{l}{with $N_{AB}^{\left(\Lambda\right)}:=\dot{C}_{AB}+\mathcal{L}_{U_{\left(0\right)}}C_{AB}-\frac{1}{2}\left(D_{C}U_{\left(0\right)}^{C}\right)C_{AB}$}\tabularnewline
					&  & \multicolumn{2}{l}{$\qquad\qquad\qquad\qquad-\frac{\Lambda}{6}\gamma_{AB}C_{CD}C^{CD}$}\tabularnewline
					&  & \multicolumn{2}{c}{}\tabularnewline
					Inputs to arrive at a & Equations of motion (asymptotic form of the & \multicolumn{2}{l}{Equations of motion (asymptotic form of the}\tabularnewline
					formula for the mass & solution should include the generic graviton) & \multicolumn{2}{l}{solution should include the generic graviton)}\tabularnewline
					and its variation &  & \multicolumn{2}{c}{}\tabularnewline
					(energy flux) & Mass should reduce to known expressions when & \multicolumn{2}{l}{Mass should reduce to known expressions when}\tabularnewline
					& there is no radiation & \multicolumn{2}{l}{there is no radiation}\tabularnewline
					&  & \multicolumn{2}{c}{}\tabularnewline
					& Energy flux should be negative or zero & \multicolumn{2}{l}{Action principle should be well-defined}\tabularnewline
			\end{tabular}\end{ruledtabular}
		}
	\end{table*}

	\section{Bondi revisited for $\Lambda > 0$}
	\label{sec:BS-coordinates}
	Although the geometry is very different for $\Lambda>0$ and $\Lambda=0$, it turns out that in the natural extension of the coordinate system used by Bondi and Sachs the formulas for energy flux, energy, and
	the like turn out to be remarkably simple, and furthermore reduce for $\Lambda=0$ to theirs. For this reason, we will go right away into
	the analysis in that particular coordinate system. 

	\subsection{Asymptotic behavior of the metric}
	
	In the coordinate system $\left(u,r,\theta,\phi\right)$ originally
	introduced by Bondi \citep{Bondi:1960jsa} and generalized later to
	the non-axisymmetric case by Sachs \citep{Sachs:1962wk,Sachs:1962zza},
	the line element reads
	\begin{align}
		ds^{2} & =e^{2\beta}\frac{V}{r}\;du^{2}-2e^{2\beta}\;dudr\nonumber \\
		& +r^{2}g_{AB}\left(dx^{A}-U^{A}\;du\right)\left(dx^{B}-U^{B}\;du\right)\; \label{eq:metric-Bondi-gauge}
	\end{align}
	with $-\infty < u < \infty$ and $0<r<\infty$.
	The $x^{A}$ are coordinates on the two-sphere, which we choose here to be the standard spherical one: $x^A=\left(\theta,\phi\right)$ with $0\leq\theta\leq\pi$ and $0\leq\phi<2\pi$.\footnote{Strictly speaking, one of course needs two charts to cover the 2-sphere.} The coordinate
	$u$ is null because when $du=0$ and $dx^{A}=0$, one has $ds^{2}=0$.
	Radiation is ``observed'' as $r\rightarrow\infty$. In this limit, one approaches the future boundary --- often denoted by $\scri$. These coordinates nicely encode that the topology of $\scri$ is $\mathbb{R} \times \mathbb{S}^2$, which is the relevant setting for studying gravitational radiation emitted by compact sources.

	The functions $\beta$, $V$, $g_{AB}$ and $U^{A}$ depend on $x^{A}$,
	$u$, and $r$. The procedure is to expand the metric components in
	powers of $r^{-1}$, demand reasonable boundary conditions and impose Einstein's equations order by order
	in $r$. The latter step does not restrict the dependence on $u$ and $x^{A}$, but leads to relationships between different coefficients in the expansion.
	We will omit the details of this calculation and state the result to the order needed
	for the determination of possible asymptotic ``charges,'' and their
	fluxes.
	
	One finds
	\begin{subequations}
		\label{eq:fall-off-physical}
		\begin{align}
			\beta & =-\frac{1}{32r^{2}}C^{AB}C_{AB} \nonumber \\
			& \;\;+\frac{1}{128r^{4}}\left(\left(C^{AB}C_{AB}\right)^{2}-12C^{AB}E_{AB}\right)+\ldots, \label{eq:fall-off-beta}\\
			V & =\frac{\Lambda r^{3}}{3}-D_{A}U_{\left(0\right)}^{A}\;r^{2}-\left(1+\frac{\Lambda}{16}C^{AB}C_{AB}\right)r\nonumber \\
			& \qquad +2M+\ldots \label{eq:fall-off-V}\\
			U^{A} & =U_{(0)}^{A}-\frac{1}{2r^{2}}D_{B}C^{AB}\nonumber \\
			& \qquad -\frac{2}{3r^{3}}\left(N^{A}-\frac{1}{2}C^{AB}D^{C}C_{BC}\right)+\ldots, \label{eq:fall-off-UA}\\
			g_{AB} & =\gamma_{AB}+\frac{C_{AB}}{r}+\frac{C^{CD}C_{CD}\gamma_{AB}}{4r^{2}}+\frac{E_{AB}}{r^{3}}+\ldots, \label{eq:fall-off-gAB} \\
			\det g_{AB} & =\sin^{2}\theta.  \label{eq:fall-off-determinant}
		\end{align}
	\end{subequations}
	These expressions depend on $\Lambda$ explicitly in Eq.~\eqref{eq:fall-off-V} and through $U_{(0)}^{A}\left(\Lambda\right)$, which vanishes for $\Lambda=0$,
	but depends implicitly on it according to Eq.~\eqref{eq:constraint-einstein} below. When $\Lambda=0$, they reduce to those of Sachs.
	Here $D_{A}$ is the covariant derivative with respect to the metric
	of the unit two-sphere $\gamma_{AB}$. The indices $A,B$ are lowered
	and raised with the metric $\gamma_{AB}$. The symmetric tensors $C_{AB}$
	and $E_{AB}$ are traceless: $\gamma^{AB}C_{AB}=\gamma^{AB}E_{AB}=0$.
	
	Besides Eqs.~\eqref{eq:fall-off-physical}, there are two further restrictions on the coefficients which are of decisive importance in the analysis. They are the following:
	(i) The zeroth order term in $g_{AB}$ is required to be the standard
	line element on the unit two-sphere:
	\begin{equation}
		\gamma_{AB}dx^{A}dx^{B}=d\theta^{2}+\sin^{2}\theta d\phi^{2}.\label{eq:ds2sphere}
	\end{equation}
	This additional demand, imposed by Bondi, which does not follow from
	Einstein's equations and it is not a mere restriction on the coordinate
	system, turns out to be of enormous consequence: It will guarantee
	later on that no divergent quantities
	appear in the analysis of a problem that has no physical singularities.
	In contradistinction, Eq.~\eqref{eq:fall-off-determinant} can be imposed to all orders by a change of coordinates $r=f\left(r',\theta,\phi\right)$.
	(ii) Besides the relations between the coefficients in Eqs.~\eqref{eq:fall-off-physical}, Einstein's equations imply 
	\begin{equation}
		2D_{(A}U_{B)}^{(0)}-\gamma_{AB}D_{C}U_{(0)}^{C}=\frac{\Lambda}{3}\;C_{AB}\;.\label{eq:constraint-einstein}
	\end{equation}
	Equation \eqref{eq:constraint-einstein} exhibits the key difference
	in the imprint of the gravitational wave on the metric for $\Lambda=0$
	versus $\Lambda\neq0$. In fact, as we will see, the tensor $C_{AB}$ describes the field of the wave when it is $u$-dependent, and we see from \eqref{eq:constraint-einstein} that
	when $\Lambda=0$, the waves do not affect the metric to the lowest order.
	However, when $\Lambda\neq0$ the wave affects the metric even to the lowest order through the shift vector $U_{\left(0\right)}^{A}$.
	
	Note that the particular solution to Eq.~\eqref{eq:constraint-einstein} exclusively exhibits modes with $\ell\geq 2$ that are inherited from the tensor $C_{AB}$. The information of the gravitational wave is exclusively contained within these modes. On the other hand, the solution of the homogeneous equation, specifically the conformal Killing equation on the 2-sphere, only has $\ell=1$ modes and are independent of the wave degrees of freedom. These latter modes represent the freedom in selecting the frame at infinity and can be set to zero without loss of generality.
	\\
	\\
	\emph{Remark.} 
	The fact that no regularization is needed at any step in the present work and that, in particular, all the charges are finite follows from allowing a generic $U^A_{(0)} \neq 0$. Had we imposed $U^A_{(0)}=0$, we would have been forced to let $\gamma_{AB}$ be a generic metric, but divergences would appear.   
	
	\subsection{Asymptotic symmetries for $\Lambda=0$ and $\Lambda>0$ compared and contrasted}
	
	\subsubsection{Mass for $\Lambda=0$}
	
	When the cosmological constant vanishes, Bondi proposed that the integral
	over a two-sphere of the coefficient $M\left(u,\theta,\phi\right)$
	appearing in Eq.~\eqref{eq:fall-off-V}
	\begin{equation}
		E=\frac{1}{4\pi G}\oint d^{2}S\;M,\label{eq:energy1}
	\end{equation}
	is the total energy of the system (with $d^{2}S=\sin\theta\,d\theta d\phi$). To validate this guess, he observed
	first that for the static Schwarzschild solution, $M$ was indeed
	the Schwarzschild mass. Then he moved on to investigate dynamical cases with gravitational waves,
	when the integral of $M$ over a large sphere was expected to diminish
	as a function of $u$ due to an energy flux emitted by a source within
	the sphere and going out to infinity (the coordinate $u$ is a retarded
	coordinate because the sign of the $dudr$ term in the line element is negative).
	This crucial test was satisfied because one can verify, from Einstein's
	equations, that
	\begin{equation}
		\frac{dE}{du}=-\frac{1}{32\pi G}\oint d^2S\,N_{AB}N^{AB}<0\ensuremath{\qquad}(\ensuremath{\Lambda=0}).\label{eq:energy-flux-flat}
	\end{equation}
	The mass expression in Eq.~\eqref{eq:energy1} has later also been derived using other methods such as the Landau-Lifschitz approach based on a pseudo-tensor (see e.g.~\cite{Thorne:1980ru}) and covariant phase space methods (see e.g.~\cite{Barnich:2011mi, Flanagan:2015pxa}).
	
	\subsubsection{Angular momentum for $\Lambda=0$}
	
	If one were to attempt guessing an expression for the angular momentum,
	one would naturally focus on the shift $N_{A}$ because it carries the
	imprint of being ``stationary'' (versus static). One would need a two-form
	to integrate over the sphere constructed out of this shift. The simplest
	candidate is its exterior derivative. So, one would write
	\begin{align}
		\vec{J} & =\frac{1}{8\pi G}\oint d^{2}S\;\hat{r}\epsilon^{AB}D_{A}N_{B}.\label{eq:angular momentum_1}
	\end{align}
	The first test would be to check if this formula gives the right
	value for the angular momentum of the Kerr-de Sitter solution (which can be
	brought to satisfy the boundary conditions in Eq.~\eqref{eq:fall-off-physical}, see Sec.~\ref{sec:Kerr}). If one does so, one finds that indeed the test
	is passed. One does not expect the angular momentum flux to have a
	definite sign so that test is not available, but a complete analysis
	of the asymptotically defined symmetries confirms its validity. The
	vector $N^{A}$ is referred to as ``angular momentum aspect''.\footnote{Beware, conventions differ on the exact definition of the angular-momentum aspect: some authors shift $N^A$ by terms proportional to $C_{AB}$ and its derivatives, and/or multiply it by a numerical factor.}
	
	\subsubsection{Symmetry for $\Lambda=0$}
	In order to prove that Eqs.~\eqref{eq:energy1} and \eqref{eq:angular momentum_1}
	are the energy and the angular momentum, one needs to show that they
	generate time translations and spatial rotations at infinity when acting on phase space. That
	proof, and much more, was given by Sachs who, in a brilliant analysis
	did two things: (i) He discovered, extending previous work of Bondi, Metzner and Van der Burg, that the asymptotically defined symmetry
	is enormously larger than the expected Poincar\'e group, and that the
	commutators of its Killing vectors form an infinite-dimensional Lie algebra now called the Bondi-Metzner-Sachs algebra \citep{Sachs:1962wk}. (ii) He \emph{postulated}
	a commutation rule for the two independent components of the news
	$C_{AB}$ and showed that, with just that, $M\left(\theta,\phi\right)$
	and the Lorentz generators $J_{\mu\nu}$ that he also constructed,
	generate the symmetry algebra \citep{Sachs:1962zza}. In particular, the zero mode \eqref{eq:energy1}
	generates time translations. By guessing the commutation rule, Sachs
	did not need to use the action principle, but just the equations of
	motion. Later developments have permitted to recover the canonical
	generators of the Bondi-Metzner-Sachs algebra from the action principle \cite{Barnich:2011mi, Henneaux:2018cst, Bunster:2018yjr}. 
	
	\subsubsection{Symmetry for $\Lambda>0$}
	
	For $\Lambda=0$, besides the energy and angular momentum, one has
	boosts $\vec{K}$ and infinitely many supertranslation generators
	$M\left(\theta,\phi\right)$, with spherical modes $\ell\geq1$. 
	The situation is dramatically different for $\Lambda\neq0$, in which case only
	$E$ and $\vec{J}$ are present. The complete asymptotic
	symmetry algebra consists just of time translations and spatial rotations,
	and \emph{the expressions for the generators are the same as for $\Lambda=0$}.
	This is why we have brought them out especially above. 
	
	\section{Regge-Teitelboim analysis of the symmetries for $\Lambda\protect\neq0$}
	\label{sec:symmetries}
	
	\subsection{Preservation of the asymptotic behavior of the metric}
	
	Since $\Lambda$ does not appear explicitly in the asymptotic form \eqref{eq:metric-Bondi-gauge}
	of the metric, the form of the asymptotic Killing vectors for $\Lambda\neq0$
	is the same as the one given by Sachs for $\Lambda=0$ (his equations
	III5-7 in \citep{Sachs:1962zza}), that is
	\begin{subequations}
		\begin{align}
			\xi^{u} & =T\left(u,x^{A}\right),\\
			\xi^{r} & =-\frac{r}{2}\left(D_{A}X^{A}+D_{A}I^{A}-U^{A}D_{A}T\right),\\
			\xi^{A} & =X^{A}\left(u,x^{A}\right)+I^{A}\left(u,r,x^{A}\right)\\
			& \text{with}\;\;I^{A}=-\left(D_{B}T\right)\int_{r}^{\infty}dr'\left(\frac{e^{2\beta}}{r^{2}}g^{AB}\right).
		\end{align}    
	\end{subequations}
	
	The preservation of Eq.~\eqref{eq:constraint-einstein} under the
	action of the asymptotic Killing vectors implies that $X^{A}$ must
	obey the following differential equation 
	\begin{equation}
		2D_{(A}X_{B)}-\gamma_{AB}D_{C}X^{C}=2 U^{(0)}_{(A} D_{B)} T - \gamma_{AB} U^C_{(0)}D_C T.\label{eq:param}
	\end{equation}
	The preservation of the fall-off of the metric also requires that the parameters $T$ and $X^{A}$ obey
	the following first order differential equations in time
	\begin{align}
		\dot{T} & = \frac{1}{2} D_{A}X^{A}-\frac{3}{2}U_{(0)}^A D_{A}T \,,\label{eq:Tdot}\\
		\dot{X}^{A} & =\dot{T} U^A_{(0)} - U^A_{(0)} U^B_{(0)} D_B T -\frac{\Lambda}{3}D^{A}T\,.\label{eq:Ydot}
	\end{align} 
	In particular, Eq.~\eqref{eq:Tdot} is obtained from the preservation of the decay of the $g_{ur}$ component, and Eq.~\eqref{eq:Ydot} from the $g_{uA}$ component.
	Eqs.~\eqref{eq:param}-\eqref{eq:Ydot} constrain the algebra to three rotations and the time translation as we will see in the next subsection.
	
	
	\subsection{Symmetry algebra}
	
	The symmetry algebra is determined from $T$ and $X^A$ satisfying Eqs.~\eqref{eq:param}-\eqref{eq:Ydot}. Eq.~\eqref{eq:param} constraints $T$ tremendously: from a generic function of $u,\theta,\phi$ to a function of $u$ only. Eq.~\eqref{eq:Tdot} then further requires that $T$ is time-independent, so that $T$ can only be a constant. Using this, we find that there are only three independent solutions for $X^A$ describing exactly the three rotations on the sphere. We will now show in detail how this comes about.
	
	To analyze Eq.~\eqref{eq:param}, it is useful to introduce $Y^{A}$ 
	\begin{align}
		Y^{A} & =X^{A}-U_{\left(0\right)}^{A}T\;,\label{eq:newparam}
	\end{align}
	which explicitly separates a ``frame rotation''
	at infinity. In which case, we get
	\begin{equation}
		2D_{(A}Y_{B)}-\gamma_{AB}D_{C}Y^{C}=-\frac{\Lambda}{3}TC_{AB}.\label{eq:param-Y}
	\end{equation}
	This equation has the same form as the one obeyed by the zero order shift (Eq.~\eqref{eq:constraint-einstein}), except for a negative sign --- which is just a matter of convention in the definition of $Y^A$ --- and the appearance of the factor $T$ on the right-hand side. 
	Decomposing $Y^A$ into vector spherical harmonics, we see that the left-hand side of Eq.~\eqref{eq:param-Y} contains no $\ell=1$ modes as these are in the kernel of the conformal Killing operator. Therefore, the right-hand side cannot contain any $\ell=1$ modes. Decomposing $T$ and $C_{AB}$ into spin-weighted spherical harmonics
	\begin{align}
		T &= \sum_{\ell,m} T_{\ell m}(u) \; {}_0Y_{\ell m}\\
		C_{AB} & = \sum_{\ell,m} C^E_{\ell m}(u) \left( {}_{-2}Y_{\ell m} m_A m_B + {}_{2}Y_{\ell m} \bar{m}_A \bar{m}_B\right) \notag \\
		& \qquad - i C^B_{\ell m}(u) \left( {}_{-2}Y_{\ell m} m_A m_B - {}_{2}Y_{\ell m} \bar{m}_A \bar{m}_B\right)
	\end{align}
	where $m_A, \bar{m}_A$ are complex null vectors on the two-sphere satisfying $m^A \bar{m}_A =1$, we find that if we project onto $\bar{m}^A \bar{m}^B$, their product can be written as
	\begin{align}
		T \bar{m}^A \bar{m}^B C_{AB} =  \sum_{\ell,m} \mathcal{C}_{\ell m} \; {}_{-2}Y_{\ell m} \; .
	\end{align}
	So we need to determine what the constraints on $T_{\ell m}$ are such that $\mathcal{C}_{\ell m}$ does not contain any $\ell=1$ modes. We find that
	\begin{align}
		\mathcal{C}_{\ell m} &= \sum_{\ell',m',\ell'',m''} T_{\ell' m'} \left(C^E_{\ell'' m''} - i C^B_{\ell'' m''} \right) \times \notag \\
		& \qquad \qquad \qquad \int d^2 S \; {}_0Y_{l'm'} \; {}_{-2}Y_{\ell'' m''} {}_{-2}\bar{Y}_{\ell m} \label{eq:c-coef}\\
		&= \sum_{\ell',m',\ell'',m''} \sqrt{\frac{(2\ell+1)(2\ell'+1)(2\ell''+1)}{4\pi}} \times \notag \\
		& \qquad \qquad \qquad (-1)^m T_{\ell' m'} \left(C^E_{\ell'' m''} - i C^B_{\ell'' m''} \right) \times \notag \\
		& \qquad \qquad \qquad
		\begin{pmatrix}
			\ell & \ell' & \ell'' \\
			m & m' & m''
		\end{pmatrix}
		\begin{pmatrix}
			\ell & \ell' & \ell'' \\
			-2 & 0 & 2
		\end{pmatrix}. \label{eq:3j-symbols}
	\end{align}
	where in going from Eq.~\eqref{eq:c-coef} to \eqref{eq:3j-symbols}, we used that ${}_{s}\bar{Y}_{\ell m}=(-1)^{m+s}{}_{-s}Y_{\ell m}$ and that the integral over three spin-weighted spherical harmonics is given by the product of two 3$j$-symbols \cite[Eq.~(34.3.22)]{NIST:DLMF}. Spin-weighted spherical harmonics ${}_sY_{\ell m}$ are not defined for $|s|>\ell$ so $C^{E}_{\ell m}/C^{B}_{\ell m}$ does not have any modes with $\ell=0$ or $1$.
	Hence, $\mathcal{C}_{\ell m}$ contains no $\ell=1$ modes only if $T_{\ell m}$ is non-zero for $\ell=0$, because $C^E_{\ell m}/C^B_{\ell m}$ is generically non-zero for $\ell \ge 2$ and the 3$j$-symbols are non-zero when $|\ell'+\ell''| \le \ell \le \ell' + \ell''$. 
	
	So far, we have seen that $T(u,\theta,\phi)=T_0(u)$ and $X^A$ satisfies the conformal Killing equation. Substituting this into Eq.~\eqref{eq:Tdot}, we obtain
	\begin{equation}
		\dot{T}_0 = \frac{1}{2}  D_A X^A . \label{eq:T0dot}
	\end{equation}
	The only consistent solution is if both sides of the equation vanish independently. Hence, we find that $T_0$ is $u$-independent and $X^A$ is also divergence-free.  Finally, from Eq.~\eqref{eq:Ydot}, we obtain that $X^A$ is time-independent. 
	Therefore, we find that $T$ and $X^A$ are 
	\begin{equation}
		T=T_{0}\qquad\text{and}\qquad X^{A}=\epsilon^{AB}D_{B}\left(\vec{\Omega}\cdot\hat{r}\right) ,\label{eq:algebra}
	\end{equation}
	for constant $T_{0}$ and $\vec{\Omega}$.
	Substituting this back into the form of the asymptotic Killing vector fields, we obtain
	\begin{subequations}
		\begin{align}
			\xi^{u} & =T_0,\\
			\xi^{r} & =0, \\
			\xi^{A} & =\epsilon^{AB}D_{B}\left(\vec{\Omega}\cdot\hat{r}\right).
		\end{align}    
	\end{subequations}
	One immediately recognizes this as the $\mathbb{R} \oplus so\left(3\right)$ algebra. This result generalizes the findings in \citep{abk1}, where it was shown that the asymptotic symmetry group is exactly the 4-dimensional group of time translations and rotations when $\scri$ has  $\mathbb{R} \times \mathbb{S}^2$ topology \emph{and} the induced metric at $\scri$ is conformally flat. The requirement of conformal flatness, which severely restricted the allowed gravitational radiation by essentially cut the degrees of freedom of the gravitational field in half, can be lifted.

	\section{Revisiting Regge-Teitelboim for $\Lambda>0$ and radiation at future infinity}
	\label{sec:charges-and-fluxes}
	
	\subsection{Charges}
	The variation of the charge is obtained using the covariant approach of Barnich and Brandt \cite{Barnich:2001jy}, which --- as they proved --- is equivalent to the standard Regge-Teitelboim analysis \citep{Regge:1974zd}.\footnote{Note that it is also equivalent to the Wald-Zoupas method \cite{Lee:1990nz,Wald:1999wa} for an appropriate choice of boundary terms (see e.g. \cite{Compere:2018aar}), as well as to the one of Abott, Deser and Tekin \cite{Abbott:1981ff, Deser:2002rt, Deser:2002jk}.}
	In particular, if $h_{\mu \nu}:=\delta g_{\mu \nu}$ corresponds to the functional variation of the spacetime metric, then the general expression for the variation of the charge is given by
	\begin{align}
		\delta_{\xi} Q&=\frac{1}{16\pi G}\oint_{\scri}\left(d^{2}x\right)_{\mu\nu} \left[\xi^{\nu}\nabla^{\mu}h-\xi^{\nu}\nabla_{\sigma}h^{\mu\sigma}+\xi_{\sigma}\nabla^{\nu}h^{\mu\sigma} \right. \notag \\
		& \left. \qquad +\frac{1}{2}h\nabla^{\nu}\xi^{\mu}+\frac{1}{2}h^{\nu\sigma}\left(\nabla^{\mu}\xi_{\sigma}-\nabla_{\sigma}\xi^{\mu}\right)-\left(\mu\leftrightarrow\nu\right)\right], \label{eq:BBcharge}
	\end{align}
	where $\xi^{\mu}$ is the asymptotic Killing vector, $h:=g^{\mu \nu}h_{\mu \nu}$ and the volume form is
	\begin{equation}
		\left(d^{2}x\right)_{\mu\nu}:=\frac{1}{4}\epsilon_{\mu\nu\lambda\sigma}dx^{\lambda}\wedge dx^{\sigma}.
	\end{equation}
	Applying this to our set-up, we find
	\begin{align}
		\delta_\xi Q & =\frac{1}{16\pi G}\oint_{\scri}d^{2}S\,\left[T\delta\left(4M\right) +\frac{T}{2}N_{AB}^{\left(\Lambda\right)}\delta C^{AB}\right.\nonumber \\
		& \left. +X^{A}\delta\left(2N_{A}+\frac{1}{16}D_{A}\left(C_{BD}C^{BD}\right)\right) \right. \nonumber \\
		& \left. - T U^{A}_{(0)}\delta\left(2N_{A}+\frac{1}{16}D_{A}\left(C_{BD}C^{BD}\right)\right) \right] \; , \label{eq:deltaQ}
	\end{align}
	where the tensor $N_{AB}^{\left(\Lambda\right)}$ is defined by 
	\begin{align}
		N_{AB}^{\left(\Lambda\right)} & :=\dot{C}_{AB}+\mathcal{L}_{U_{\left(0\right)}}C_{AB}-\frac{1}{2}\left(D_{C}U_{\left(0\right)}^{C}\right)C_{AB}\nonumber \\
		& -\frac{\Lambda}{6}\gamma_{AB}C_{CD}C^{CD}.\label{eq:News}
	\end{align}
	$N_{AB}^{(\Lambda)}$ generalizes the Bondi News tensor when the cosmological constant is non-zero.
	This expression
	acquires a similar structure as the one obtained in \citep{Barnich:2011mi}
	for the asymptotically flat case with $M$ playing the role of the
	``Bondi mass aspect'' and $N^{A}$ that of the ``angular momentum
	aspect''. However, there are some differences that come from the
	presence of a non-zero cosmological constant. Apart from the correction
	coming from $\Lambda$ in the ``News tensor'' $N_{AB}^{(\Lambda)}$
	in Eq.~\eqref{eq:News}, there is an additional non-integrable term proportional to $U^A_{(0)}$ that vanishes in the limit when $\Lambda\rightarrow0$.
	
	It is worth emphasizing that the variation of the charge is finite
	in the limit when $r\rightarrow\infty$, \emph{without the need of
		any ad-hoc regularization procedure}. The only potentially divergent
	terms were those proportional to $r$, which after some appropriate
	integration by parts on the sphere acquire the form 
	\begin{align}
		\delta_\xi Q_{\text{div}} & =-\frac{r}{32\pi G}\oint d^{2}S\Big( 2D_{(A}X_{B)}-\gamma_{AB}D_{C}X^{C} \notag \\
		&  -2 U_{(A}^{(0)} D_{B)}T + \gamma_{AB} U^C_{(0)}D_C T  \notag \\
		&  - T \left[2  D_{(A} U^{(0)}_{B)} - \gamma_{AB}  D_C U^C_{(0)}-\frac{\Lambda}{3}C_{AB} \right]\Big)\delta C^{AB},
	\end{align}
	which, by virtue of Eqs.~\eqref{eq:constraint-einstein} and ~\eqref{eq:param}, vanishes identically. 
	
	Thus, if one assumes that $\delta T_{0}=\delta\vec{\Omega}=0$, then
	the integrable part (in the functional sense) of the variation of the charge, takes the form 
	\begin{align}
		Q^{\text{int}} [ T_0, \vec{\Omega} ] &=T_{0} \; E +\vec{\Omega}\cdot \vec{J}, \label{Qint}
	\end{align}
	where the energy $E$ and angular momentum $\vec{J}$ are
	\begin{align}
		E & =\frac{1}{4\pi G}\oint d^{2}S\;M,\label{Energy}\\
		\vec{J} & =\frac{1}{8\pi G}\oint d^{2}S\;\hat{r}\epsilon^{AB} D_{A}N_{B}.\label{Angularmomentum}
	\end{align}
	Note that the term proportional to $T$ in the last line of Eq.~\eqref{eq:deltaQ} does not contribute to the mass, because it does not contain any $\ell=0$ modes (this can again be seen from an analysis of the $3j$-symbols and noting that $U_A^{(0)}$ is only non-zero for  $\ell\ge 2$).
	As we will show in Sec.~\ref{sec:Kerr}, these expressions give the expected results for the mass and angular momentum for the Kerr-de
	Sitter geometry, and allows to extend to notion of energy and angular momentum to the case when gravitational waves are present.

	\subsection{Fluxes}
	
	The fluxes of energy and angular momentum can be directly obtained
	by taking the time derivative of Eqs.~\eqref{Energy} and \eqref{Angularmomentum}
	in conjunction with Einstein's equation. In particular, Einstein's equations yield the evolution of $M$ and $N^{A}$, respectively. The resulting expressions are rather long but manageable:
	\begin{align}
		\dot{M} & =\frac{1}{4}D_{A}D_{B}N_{(\Lambda)}^{AB}-\frac{1}{8}N_{AB}^{(\Lambda)}N_{(\Lambda)}^{AB}+\frac{\Lambda}{96}C^{AB} D^2 C_{AB}\nonumber \\
		& -\frac{\Lambda}{12}C^{AB}C_{AB}-\frac{\Lambda}{6}D_{A}N^{A}-\frac{\Lambda}{96}\left(D_{C}C_{AB}\right)\left(D^{C}C^{AB}\right)\nonumber \\
		& +\frac{\Lambda^{2}}{24}C^{AB}E_{AB}-\frac{7\Lambda^{2}}{1152}\left(C^{AB}C_{AB}\right)^{2}\nonumber \\
		& -U_{(0)}^{A}D_{A}M-\frac{3}{2}MD_{A}U_{(0)}^{A}-\frac{1}{8}C^{AB}D_{A}D_{B}D_{C}U_{(0)}^{C}\label{eq:M-dot}
	\end{align}
	and
	\begin{align}
		\dot{N}^{A} & =D^{A}M+\frac{1}{4}D^{A}D^{B}D^{C}C_{BC}-\frac{1}{4}D_{B} D^2 C^{AB}\nonumber \\
		& +\frac{5}{16}C^{AB}D^{C}N_{BC}^{(\Lambda)}-\frac{3}{16}C_{BC}D^{B}N_{(\Lambda)}^{AC}-\frac{\Lambda}{2}D_{B}E^{AB}\nonumber \\
		& -\frac{1}{2}N_{(\Lambda)}^{AB}D^{C}C_{BC}+\frac{1}{16}N_{(\Lambda)}^{BC}D^{A}C_{BC}+D_{B}C^{AB}\nonumber \\
		& +\frac{5\Lambda}{32}C_{BD}C^{CD}D_{C}C^{AB}+\frac{7\Lambda}{48}C^{AB}C^{CD}D_{B}C_{CD}\nonumber \\
		& -U_{(0)}^{B}D_{B}N^{A}+N^{B}D_{B}U_{(0)}^{A}-2N^{A}D_{C}U_{(0)}^{C}\nonumber \\
		& -\frac{1}{64}U_{(0)}^{A}\left(C_{BD}C^{BD}\right)-\frac{1}{64}\left(D^2 U_{(0)}^{A}\right)C_{BD}C^{BD}\nonumber \\
		& +\frac{1}{32}D^{A}\left(D_{C}U_{(0)}^{C}\right)\left(C_{BD}C^{BD}\right)\; . \label{eq:NA-dot}
	\end{align}
	The energy flux is given by 
	\begin{align}
		\frac{dE}{du} & =-\frac{1}{32\pi G}\oint d^{2}S\left\{ N_{AB}^{(\Lambda)}N^{AB}_{(\Lambda)}+\frac{2\Lambda}{3}C^{AB}C_{AB}\right.\nonumber \\
		& -\frac{\Lambda}{6}C^{AB} D^2 C_{AB}+\frac{7\Lambda^{2}}{144}\left(C^{AB}C_{AB}\right)^{2}-\frac{\Lambda^{2}}{3}C^{AB}E_{AB}\nonumber \\
		& \left.+\left(4M+D_{A}D_{B}C^{AB}\right)\left(D_{C}U_{\left(0\right)}^{C}\right)\right\} .\label{eq:radrate}
	\end{align}
	The first term on the right-hand side has the same form as the one
	that contributes to the loss of energy in the asymptotically flat
	case. However, there are now also corrections coming from the presence
	of the cosmological constant which are up to fourth order in the
	fields. These higher order terms are characteristic of the full nonlinear
	theory and cannot be seen in the linearized approximation. In Sec.~\ref{sec:linearized-charges-fluxes},
	we will show that when the higher order terms are neglected, the total
	amount of energy radiated in a certain interval of time precisely
	coincide with the one reported in \citep{abk2,Chrusciel:2020rlz,Kolanowski:2020wfg}.
	An important difference with the asymptotically flat case is that
	the flux of energy is not manifestly negative. This was also observed
	for the case of homogeneous gravitational perturbations on a de Sitter
	background in \citep{abk2}. Moreover, this can also occur for Maxwell
	fields on a de Sitter background \citep{abk2}
	and thus seems a rather generic feature of spacetimes with $\Lambda>0$.
	This is likely due to the fact that there is no global time-like Killing
	vector field in de Sitter spacetime. However, as was pointed out in
	\citep{abk3}, and as we will show in Sec.~\ref{sec:example-linear-waves},
	in the case of quadrupolar radiation in the linearized theory, the
	flux of energy \emph{is} manifestly negative.
	
	Analogously, the flux of angular momentum takes the form 
	\begin{equation}
		\frac{d\vec{J}}{du}=\frac{1}{8\pi G}\oint d^{2}S\;\hat{r} \; \epsilon^{AB}D_{A}\dot{N}_{B},\label{rateangular}
	\end{equation}
	where $\dot{N}_{B}$ is given by Eq.~\eqref{eq:NA-dot}. Due to the cosmological constant there is no angular momentum ambiguity, because there are no abelian supertranslations
	as is the case with $\Lambda=0$.
	
	The flux of energy and angular momentum
	in Eqs.~\eqref{eq:radrate}-\eqref{rateangular} can alternatively
	be obtained from the non-integrable part of the variation of the charge
	in Eq.~\eqref{eq:deltaQ} following the prescription in \citep{Barnich:2011mi}
	(see also \citep{Bunster:2018yjr,Bunster:2019mup}). We have verified this explicitly for the energy flux.

	\subsection{No radiation condition}
	In the Bondi-Sachs coordinates we introduced in Sec.~\ref{sec:BS-coordinates}, the presence of a gravitational wave manifests itself through the tensor $C_{AB}$ (and, of course, $U^A_{(0)}$ given its direct link to $C_{AB}$ through Eq.~\eqref{eq:constraint-einstein}).
	In particular, whenever $C_{AB}$ is time-dependent, there is gravitational radiation. This is the case for all the examples in Sec.~\ref{sec:examples}.
	This interpretation is further supported by the expressions for flux of energy and angular momentum in Eq.~\eqref{eq:radrate} and Eq.~\eqref{rateangular}.
	Whenever the spacetime under consideration is stationary, these fluxes vanish trivially as $\dot{M}$ and $\dot{N}_A$ are both zero. 
	When $C_{AB}$ is zero, the flux formulas are also both trivially zero.  
	The scenario in which $C_{AB}$ is non-zero, but time-independent is subtle. From the expressions for the fluxes, it is not evident that they will both be zero. Although we have no formal proof, we have some evidence that this is indeed the case. First, we have two non-trivial examples in which this expectation is borne out. In particular, time-independent linearized quadrupolar waves and linearized Robinson-Trautman solutions. We will discuss the general time-dependent cases in detail in Sec.~\ref{sec:example-linear-waves} and Sec.~\ref{sec:RT-solution}, respectively. When we restrict those solutions to $C_{AB}$ non-zero but $u$-independent, the integrand for the energy flux is non-zero in both cases, but the resulting integral vanishes (specifically, the first two lines cancel with the last line in Eq.~\eqref{eq:radrate}). 
	This is evident from the final expressions in Eqs.~\eqref{eq:flux-energy-lin} and \eqref{eq:flux-RT-lin}, which vanish if $\partial_u C_{AB} = 0$. The cancellation is  non-trivial and all terms conspire for this to happen. 
	The second line of evidence is by analogy with a related scenario in the asymptotically flat case. For asymptotically flat spacetimes, the presence of radiation also manifests itself through a $u$-dependent $C_{AB}$, or equivalently, a non-zero Bondi News tensor $N_{AB} = \partial_u C_{AB}$. 
	Thus, if $C_{AB}(u,\theta,\phi)= u \bar{C}_{AB} (\theta,\phi)$, $N_{AB}$ is non-zero and one would classify this spacetime as radiative. However, the resulting energy flux in this case would be constant as $N_{AB}$ is time-independent and the flux only depends on $N_{AB}$. As a result, the total energy radiated would be infinite. This is unphysical and therefore not considered to be a viable solution, and one typically (implicitly) does not include such solutions. We expect a similar scenario to hold here: If $C_{AB}$ is a non-zero, but $u$-independent solution, either the flux will be zero, or it will be constant and as a result the energy radiated will be infinite. The latter is not physical and we do not include such solutions in our solution space.

	\section{Application to special cases}
	\label{sec:examples}
	In this section, we show explicitly that the fall off conditions in
	Eq.~\eqref{eq:fall-off-physical} accommodate a wide range of physically
	interesting solutions to Einstein's equation. First, we discuss the de Sitter spacetime itself before moving on to two
	black hole solutions in the presence of a positive cosmological constant:
	the non-rotating Schwarzschild-de Sitter spacetime and the rotating Kerr-de Sitter spacetime. Next, we discuss linearized solutions to
	Einstein's equations with $\Lambda>0$ representing gravitational radiation emitted by a compact source. Finally, we describe a simple model of gravitational radiation with a single degree of freedom known
	as the Robinson-Trautman spacetime.
	
	\subsection{de Sitter spacetime}
	\label{sec:example-de-sitter}
\begin{figure}
		\includegraphics[width=0.25\textwidth]{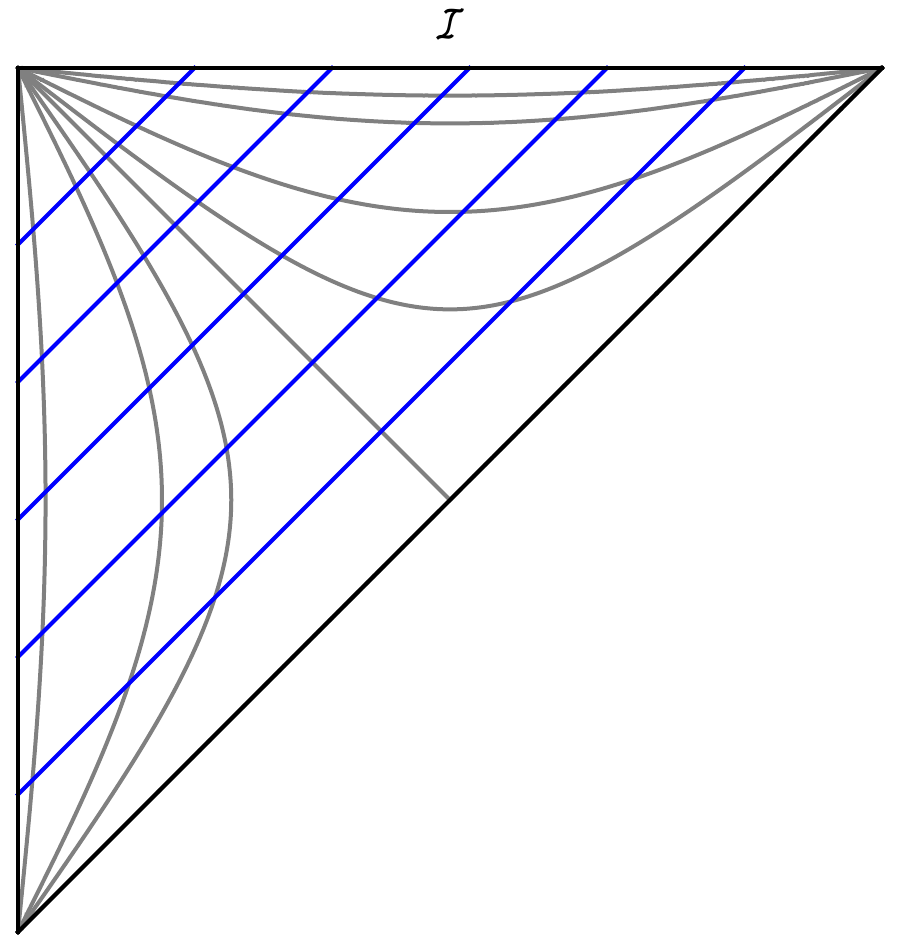}
		\caption{
			\label{fig:conformal-diagram-dS}
			Conformal diagram of de Sitter spacetime, where each point represents a two-sphere (except for the points at $r=0$ which are proper points). The gray lines denote surfaces of constant $r$, and the blue lines surfaces of constant $u$ (see Eq.~\ref{eq:exact-de-Sitter}). $r=0$ is the vertical line on the left, and $r\to \infty$ at $\scri$. The diagonal constant $r$ surface corresponds to the cosmological horizon: $r= \ell$.} 
	\end{figure}

	The full de Sitter spacetime is not an example of the class of spacetimes we have defined. This is not problematic, as the goal of this paper is to describe radiation generated by compact sources in the presence of $\Lambda$ in which case not the complete de Sitter spacetime, but the Poincar\'e
	patch of de Sitter spacetime with an additional point at $\scri$ removed is relevant. The removal of this additional point is natural as it represents the intersection of the future boundary with the source generating radiation (see also \cite[Sec. II]{abk3}). As a result, the future boundary has topology
	$\mathbb{R}\times\mathbb{S}^{2}$ and is naturally coordinatized by $(u,r,\theta,\phi)$:
	\begin{equation}
		ds^{2} =-\left(1-\frac{\Lambda r^{2}}{3}\right)du^{2}-2dudr +r^{2}\left(d\theta^{2}+\sin^{2}\theta d\phi^{2}\right)\;. \label{eq:exact-de-Sitter}
	\end{equation}
	The behavior of these coordinates is illustrated in the conformal diagram in Fig.~\ref{fig:conformal-diagram-dS}.
	The time translation vector field $\frac{\partial}{\partial u}$ and the three rotational Killing vector fields are not only asymptotic symmetries, but symmetries of the entire spacetime. Translations and inverted translations, which are symmetries of the full de Sitter spacetime, do not leave $i^0$ and $i^+$ invariant and are therefore not permissible (for a more extensive discussion, see \citep{abk1}). 
	
	\subsection{Schwarzschild-de Sitter spacetime}
	The simplest prototype for describing non-dynamical isolated gravitating
	systems in the presence of a cosmological constant is undoubtedly
	the Schwarzschild-de Sitter spacetime. This spacetime describes a
	non-rotating black hole with $\Lambda>0$. We consider the metric
	in Eddington-Finkelstein coordinates $(u,r,\theta,\phi)$: 
	\begin{align}
		ds^{2} & =-\left(1-\frac{2m}{r}-\frac{r^{2}}{l^{2}}\right)du^{2}-2dudr\nonumber \\
		& \;\;+r^{2}\left(d\theta^{2}+\sin^{2}\theta d\phi^{2}\right)\;,
	\end{align}
	where $\Lambda=3l^{-2}$. The coordinate ranges are $u\in(-\infty,\infty),r\in(0,\infty),\theta\in\left[0,\pi\right)$
	and $\phi\in\left[0,2\pi\right)$. While these coordinates do not
	provide a global chart of the spacetime, they suffice to cover the
	asymptotic region near $\scri$. In terms of the asymptotic expansions
	of the metric in Eq.~\eqref{eq:fall-off-physical}, this metric has
	$M(u,\theta,\phi)=m$ and all other coefficients zero. In particular,
	$C_{AB}$ and $U_{(0)}^{A}$ are both zero so that there is no gravitational radiation.
	
	This metric has four Killing vector fields: one Killing vector field
	generates time translations and the other three describe the spherical
	symmetry of the spacetime. A well-known
	property of (global) Killing vector fields is that every Killing field
	of the physical spacetime admits an extension to the boundary and
	is tangential to it. This is also the case for the
	above Killing vector fields, which coincide
	exactly with the asymptotic symmetry vector fields. All charges and fluxes vanish
	except for the mass, which is 
	\begin{equation}
		E_{{\rm Sch-dS}}=\frac{m}{G}\;.
	\end{equation}
	

	\subsection{Kerr-de Sitter spacetime}
	\label{sec:Kerr} 
	Kerr-de Sitter spacetimes are stationary, vacuum
	solutions to Einstein's equations describing rotating black holes
	in the presence of a positive cosmological constant. Let us consider
	the Kerr de-Sitter metric in standard Boyer-Lindquist coordinates
	$(T,R,\Theta,\Phi)$ \citep{Carter:1968ks} 
	\begin{align}
		ds^{2} & =-2a\sin^{2}\Theta\left(\frac{2mR}{a^{2}\cos^{2}\Theta+R^{2}}+\frac{a^{2}+R^{2}}{l^{2}}\right)dTd\Phi \nonumber \\
		& -\left(1-\frac{2mR}{a^{2}\cos^{2}\Theta+R^{2}}-\frac{a^{2}\sin^{2}\Theta+R^{2}}{l^{2}}\right)dT{}^{2} \nonumber \\
		& +\sin^{2}\Theta\left(\frac{2a^{2}mR\sin^{2}\Theta}{a^{2}\cos^{2}\Theta+R^{2}}+\left(a^{2}+R^{2}\right)\left(1+\frac{a^{2}}{l^{2}}\right)\right)d\Phi^{2} \nonumber  \\
		& +\left(a^{2}\cos^{2}\Theta+R^{2}\right)\left(\frac{dR^{2}}{R^{2}-\left(a^{2}+R^{2}\right)\frac{R^{2}}{l^{2}}-2mR+a^{2}}\right. \nonumber \\
		& \left.+\frac{d\Theta^{2}}{1+\frac{a^{2}\cos^{2}\Theta}{l^{2}}}\right),
	\end{align}
	where the parameter $a$ is related to the amount of rotation of this
	rotating black hole. 
	In the limit, $a\to0$ one recovers the Schwarschild-de Sitter metric
	in static coordinates. Note that these Boyer-Lindquist coordinates
	are `twisted' at $\scri$: for instance, surfaces of constant $T,R$
	describe deformed spheres (consequently, the range of $\Theta,\Phi$
	is not the standard range for coordinates on the sphere). Inspired
	by the coordinate transformation used in \citep{Henneaux:1985tv}
	to undo this twisting, we perform the following asymptotic change
	of coordinates\footnote{The Kerr-de Sitter solution in Bondi coordinates was also written in \citep{Hoque:2021nti}.}
	\begin{align*}
		T & =u+l\,\text{arctanh}\left(\frac{r}{l}\right)-\frac{ml^{4}\left(1-\frac{a^{2}\sin^{2}\theta}{2l^{2}}\right)}{2\left(1+\frac{a^{2}\sin^{2}\theta}{l^{2}}\right)^{5/2}}\frac{1}{r^{4}}+\ldots\\
		R & =r\sqrt{1+\frac{a^{2}\sin^{2}\theta}{l^{2}}}-\frac{\left(1+\frac{a^{2}}{l^{2}}\right)a^{2}\sin^{2}\theta}{2\left(1+\frac{a^{2}\sin^{2}\theta}{l^{2}}\right)^{3/2}}\frac{1}{r}\\
		& -\frac{ma^{2}\sin^{2}\theta}{2\left(1+\frac{a^{2}\sin^{2}\theta}{l^{2}}\right)^{2}}\frac{1}{r^{2}}+\ldots\\
		\Theta & =\arccos\left(\frac{\cos\theta}{\sqrt{1+\frac{a^{2}\sin^{2}\theta}{l^{2}}}}\right)-\frac{a^{2}\sin\left(2\theta\right)\sqrt{1+\frac{a^{2}}{l^{2}}}}{4\left(1+\frac{a^{2}\sin^{2}\theta}{l^{2}}\right)^{2}}\frac{1}{r^{2}}\\
		& +\frac{3a^{4}\sin\left(2\theta\right)\left(1-2\sin^{2}\theta\left(1+\frac{a^{2}}{2l^{2}}\right)\right)\sqrt{1+\frac{a^{2}}{l^{2}}}}{16\left(1+\frac{a^{2}\sin^{2}\theta}{l^{2}}\right)^{4}}\frac{1}{r^{4}}+\ldots
	\end{align*}
	\begin{align*}
		\Phi & =\frac{1}{1+\frac{a^{2}}{l^{2}}}\left(\phi+\frac{a\left(u+l\text{arctanh}\left(\frac{r}{l}\right)\right)}{l^{2}}\right)\\
		& +\frac{ma^{3}\sin^{2}\theta}{4\left(1+\frac{a^{2}}{l^{2}}\right)\left(1+\frac{a^{2}\sin^{2}\theta}{l^{2}}\right)^{5/2}}\frac{1}{r^{4}}+\ldots.
	\end{align*}
	The leading terms of the Kerr-de Sitter metric near $\scri$ fit within
	our asymptotic conditions.~\footnote{Note that the solution is not in the Bondi gauge everywhere, but its asymptotic form to the orders needed is. Indeed $g_{rr}=O\left(r^{-6}\right)$ and
		$g_{rA}=O\left(r^{-4}\right)$.} The metric on $\scri$ with $u$=constant is the unit two-sphere
	with $\theta,\phi$ having their standard range, i.e., $\theta\in\left[0,\pi\right),\phi\in\left[0,2\pi\right)$
	. Moreover, $U_{\left(0\right)}^{A}$ and $C_{AB}$ are both equal
	to zero, which is consistent with the fact that there is no gravitational
	radiation in this spacetime. The mass
	and angular momentum aspect are given by: 
	\begin{align}
		M & =m\frac{\left(1-\frac{a^{2}\sin^{2}\theta}{2l^{2}}\right)}{\left(1+\frac{a^{2}\sin^{2}\theta}{l^{2}}\right)^{5/2}}\;,\\
		N^{\theta} & =0\;,\\
		N^{\phi} & =-\frac{3am}{\left(1+\frac{a^{2}\sin^{2}\theta}{l^{2}}\right)^{5/2}}.
	\end{align}
	We also find that $E_{AB}$ is 
	\begin{align}
		E_{\theta\theta} & =-\frac{ma^{2}\sin^{2}\theta}{\left(1+\frac{a^{2}\sin^{2}\theta}{l^{2}}\right)^{5/2}}\;,\\
		E_{\phi\phi} & =\frac{ma^{2}\sin^{4}\theta}{\left(1+\frac{a^{2}\sin^{2}\theta}{l^{2}}\right)^{5/2}}\;,\\
		E_{\theta\phi} & =0\;,
	\end{align}
	so that $E_{AB}$ is traceless with respect to the unit two-sphere
	metric, as it should.
	
	The mass and the angular momentum can be directly computed from the
	expressions for the charges in Eqs.~\eqref{Energy} and \eqref{Angularmomentum}
	(which also define the normalization of the Killing vectors here).
	They are given by 
	\begin{equation}
		E_{{\rm {Kerr-dS}}}=\frac{m}{G}\left(1+\frac{a^{2}}{l^{2}}\right)^{-2},\qquad J_{z}=-a\;E_{{\rm {Kerr-dS}}}.
	\end{equation}
	These results coincide with the charges obtained using Hamiltonian
	methods by Marolf and Kelly in \citep{Kelly:2012zc}.\footnote{These final expressions also agree with the gravitational charges
		defined in terms of the electric part of the Weyl tensor in \citep{abk1}
		despite the fact that the mass and angular momentum there refer to
		a differently normed Killing vector field. This is due to the fact
		that in \citep{abk1}, the $\Theta,\Phi$ coordinates were assumed
		to have the standard range on the two-sphere, which is not the case.
		If this is corrected, the results here and in \citep{abk1}
		differ exactly by the expected scaling with the Killing vector field.} Moreover, these expression also precisely coincide with the ones
	obtained for Kerr-\emph{anti}-de Sitter spacetimes after replacing
	$l\rightarrow il$ \citep{Henneaux:1985tv}. Since this spacetime
	is stationary, the fluxes are trivially zero, which we verified by
	direct computation.
	

	\subsection{Linearized solutions in de Sitter spacetime}
	\label{sec:example-linear-waves}

	\subsubsection{Linearized charges and fluxes}
	\label{sec:linearized-charges-fluxes} 
	The expressions for the charges
	and fluxes simplify drastically in the linearized context. Here, we
	will briefly comment on the linearized setting and explicitly connect
	the resulting flux of energy radiated across $\scri$ to existing
	results in the literature.
	
	Let us consider the linearized gravitational field in retarded null coordinates
	$\left(u,r,x^{A}\right)$ around the de Sitter background metric
	\begin{equation}
		d\bar{s}^{2}=-\left(1-\frac{\Lambda r^{2}}{3}\right)du^{2}-2dudr+r^{2}\left(d\theta^{2}+\sin^{2}d\phi^{2}\right).\label{eq:metric-de-sitter}
	\end{equation}
	The spacetime metric is then written as 
	\begin{equation}
		g_{\mu\nu}=\bar{g}_{\mu\nu}+ h_{\mu\nu} \; . \label{eq:split-metric-linear}
	\end{equation}
	where $h_{\mu \nu}$ is kept only up to first order. Quantities that have dimensions of length are compared with an external fixed length scale. The linearization expression \eqref{eq:split-metric-linear} is valid everywhere, not just asymptotically. 
	The fall-off of the metric in the linearized theory can be directly obtained from our asymptotic
	conditions in Eq.~\eqref{eq:fall-off-physical} by neglecting the
	terms that are quadratic in the fields. The asymptotic
	form of the metric then reads 
	\begin{subequations} 
		\label{eq:fall-off-linearized}
		\begin{align}
			\beta & =\mathcal{O}\left(\frac{1}{r^{5}}\right),\\
			V & =\frac{\Lambda r^{3}}{3}-D_{A}U_{\left(0\right)}^{A}\;r^{2}-r+2M+\ldots,\\
			U^{A} & =\left(U_{(0)}^{A}-\frac{D_{B}C^{AB}}{2r^{2}}-\frac{2N^{A}}{3r^{3}}+\ldots\right),\\
			g_{AB} & =\gamma_{AB}+ \left(\frac{C_{AB}}{r}+\frac{E_{AB}}{r^{3}}+\ldots\right),
		\end{align}
	\end{subequations} where $U_{(0)}^{A}$ obeys Eq.~\eqref{eq:constraint-einstein}.
	Note that $U_A^{(0)}$, which is of order zero in the asymptotic expansion, in which the reference length is $r$, becomes of first order in the linearized theory, in which the reference length is a fixed distance. So, both expansions do not coincide at large distances.
	The time derivatives of $M$ and $N^{A}$ reduce to
	\begin{align}
		\dot{M} & =\frac{1}{2}D_{A}D_{B}N_{(\Lambda)}^{AB}-\frac{\Lambda}{6}D_{A}N^{A},\label{eq:M-dot-1}\\
		\dot{N}^{A} & =D^{A}M+\frac{1}{4}D^{A}D^{B}D^{C}C_{BC}-\frac{1}{4}D_{B} D^2 C^{AB}\nonumber \\
		& \;\;-\frac{\Lambda}{2}D_{B}E^{AB}+D_{B}C^{AB},\label{eq:NA-dot-1}
	\end{align}
	while the linearized version of the News tensor now takes the form
	\begin{equation}
		N_{AB}^{(\Lambda)}=\dot{C}_{AB}.
	\end{equation}
	
	In the linearized limit, the symmetry algebra of the de Sitter background metric can be naturally used (see Sec.~\ref{sec:example-de-sitter}).
	The radiation rates in the linear theory can be obtained from the corresponding expressions in the non-linear theory by dropping the
	cubic and the quartic terms in the fields. In the case of the energy,
	we have
	\begin{align}
		\frac{dE}{du} & =-\frac{1}{32\pi G}\oint d^{2}S\left\{ N_{AB}^{(\Lambda)}N^{(\Lambda)AB}+\frac{2\Lambda}{3}C^{AB}C_{AB}\right.\nonumber \\
		& -\frac{\Lambda}{6}C^{AB} D^2 C_{AB}-\frac{\Lambda^{2}}{3}C^{AB}E_{AB}\nonumber \\
		& \left.+\left(4M+D_{A}D_{B}C^{AB}\right)\left(D_{C}U_{\left(0\right)}^{C}\right)\right\} .\label{eq:Energy-linearized}
	\end{align}
	Remarkably, if one calculates the total flux of energy radiated in
	a finite interval of time $\Delta u$ (assuming appropriate fall-off
	near the edges of $\Delta u$), one obtains perfect agreement with
	the results obtained independently by Chr\'usciel, Hoque, and Smolka
	in~\citep{Chrusciel:2020rlz}, and by Kolanowski and Lewandowski
	in~\citep{Kolanowski:2020wfg}. It also coincides with the expression
	found by Ashtekar, Bonga and Kesavan in a different set of coordinates
	\citep{abk2}. This can be seen as follows: If we re-express $C_{AB}$
	in the terms with explicit $\Lambda$ dependence by using Eq.~\eqref{eq:constraint-einstein}, we obtain 
	\begin{align*}
		\frac{dE}{du} & =-\frac{1}{32\pi G}\oint d^{2}S\left(\dot{C}_{AB}\dot{C}^{AB}-2\Lambda D^{B}U_{\left(0\right)}^{A}E_{AB}\right.\\
		& +4D^{B}U_{\left(0\right)}^{A}C_{AB}-D^{B}U_{\left(0\right)}^{A} D^2 C_{AB}\\
		& \left.+\left(4M+D_{A}D_{B}C^{AB}\right)\left(D_{C}U_{\left(0\right)}^{C}\right)\right).
	\end{align*}
	After an integration by parts on the sphere, one obtains 
	\begin{align*}
		\frac{dE}{du} & =-\frac{1}{32\pi G}\oint d^{2}S\left(\dot{C}_{AB}\dot{C}^{AB}+U_{\left(0\right)}^{A}\left[2\Lambda D^{B}E_{AB}\right.\right.\\
		& \left.\left.-4D^{B}C_{AB}+D^{B} D^2 C_{AB}-D_{A}\left(4M+D_{B}D_{C}C^{BC}\right)\right]\right).
	\end{align*}
	Using the linearized equation of motion for $N^{A}$ in \eqref{eq:NA-dot-1},
	we can write this compactly as 
	\[
	\frac{dE}{du}=\frac{1}{8\pi G}\oint d^{2}S\left(-\frac{1}{4}\dot{C}_{AB}\dot{C}^{AB}+U_{\left(0\right)}^{A}\dot{N}_{A}\right)\,.
	\]
	So that the total amount of energy radiated in the interval of time
	$\Delta u$ is given by 
	\[
	\left.E\right|_{\Delta u}=\frac{1}{8\pi G}\int_{\Delta u}du\oint d^{2}S\left(-\frac{1}{4}\dot{C}_{AB}\dot{C}^{AB}+U_{\left(0\right)}^{A}\dot{N}_{A}\right).
	\]
	Assuming that there is no flux of radiation outside this interval
	of time, we can integrate by parts in the null time and discard the
	corresponding boundary terms, so that we can write: 
	\[
	\left.E\right|_{\Delta u}=-\frac{1}{8\pi G}\int_{\Delta u}du\oint d^{2}S\left(\frac{1}{4}\dot{C}_{AB}\dot{C}^{AB}+\dot{U}_{\left(0\right)}^{A}N_{A}\right).
	\]
	Now, by virtue of our asymptotic conditions, 
	\[
	h_{uA}=-\frac{1}{r^{2}}U_{A}.
	\]
	Therefore, if we consider an asymptotic expansion of the form 
	\[
	h_{uA}=h_{uA}^{\left(2\right)}r^{2}+h_{uA}^{\left(1\right)}r+h_{uA}^{\left(0\right)}+\frac{h_{uA}^{\left(-1\right)}}{r}+\dots,
	\]
	we obtain 
	\[
	h_{uA}^{\left(2\right)}=-U_{\left(0\right)A}\qquad,\qquad h_{uA}^{\left(-1\right)}=\frac{2}{3}N_{A}.
	\]
	Thus, in terms of these variables, the total amount of energy radiated
	in the interval of time $\Delta u$ acquires the following form
	\[
	\left.E\right|_{\Delta u}=-\frac{1}{32\pi G}\int_{\Delta u}du\oint d^{2}S\left(\dot{C}_{AB}\dot{C}^{AB}-6\dot{h}_{uA}^{\left(2\right)}h_{u}^{\left(-1\right)A}\right).
	\]
	This expression precisely coincides with the ones found in~\citep{Chrusciel:2020rlz}
	and~\citep{Kolanowski:2020wfg}.
	
	
	\subsubsection{Explicit solutions for quadrupolar modes}
	Using the set-up in the previous subsection, we will now study explicit solutions to the linearized Einstein's equation. The strategy is to solve for the homogeneous solution of Einstein's equation, which corresponds in a partial wave expansion to the gravitational field away from the source generating the gravitational waves (which is assumed to be bounded).
	In particular, the homogeneous solutions corresponding to a fixed
	$\ell$ in the spherical harmonic expansion, should necessarily be
	generated by a source with multipole moment $\ell$. Even though the homogeneous solution is only valid outside the source, in
	principle we could match this solution with the ``inner'' solution
	using matched asymptotic expansions \citep{Burke:1969zz}. The matching
	with an interior solution is beyond the scope of this paper, which focuses on the solution far away from the source.
	The waves here could be considered a generalization of the ``Teukolsky waves'' from flat spacetime to de Sitter spacetime --- apart from the different gauge choice that we implement \cite{Teukolsky:1982nz}.
	
	Since the background spacetime is spherically symmetric, it is convenient
	to use similar techniques as Regge and Wheeler did when solving for
	linearized perturbations off Schwarzschild (although we will not implement the Regge-Wheeler gauge) \citep{Regge:1957td}. In particular, we will
	use separation of variables and for the angular part of the perturbations,
	we introduce scalar, vector and tensor spherical harmonics. 
	Following the notation in \citep{Martel-Poisson},
	we note that the scalar harmonics are the usual spherical-harmonic
	functions $Y_{\ell m}(x^{A})$ satisfying the eigenvalue equation
	$D^2 Y_{\ell m}=-\ell\left(\ell+1\right)Y_{\ell m}$. There are
	two types of vector harmonics: even-parity $Y_{A}^{\ell m}$ (also known as
	electric) and odd-parity $X_{A}^{\ell m}$ (also known as magnetic), which
	are related to the scalar harmonics through the covariant derivative
	operator compatible with $\gamma_{AB}$ 
	\begin{align}
		Y_{A}^{\ell m} & :=D_{A}Y_{\ell m}\\
		X_{A}^{\ell m} & := - \epsilon_{A}^{\;\;B}D_{B}Y_{\ell m}\;.
	\end{align}
	The even- and odd-parity harmonics are orthogonal in the sense that
	$\int d^{2}S\;\bar{Y}_{\ell m}^{A}X_{A}^{\ell'm'}=0$. The tensor
	harmonics also come in the same two types: 
	\begin{align}
		Y_{AB}^{\ell m} & :=D_{(A}Y_{B)}^{\ell m}-\frac{1}{2}\gamma_{AB}D_{C}Y_{\ell m}^{C}\\
		X_{AB}^{\ell m} & :=-\epsilon_{(A}^{\;\;\;\;C}D_{B)}Y_{C}^{\ell m}\;.
	\end{align}
	These operators are traceless, i.e., $\gamma^{AB}Y_{AB}^{\ell m}=0=\gamma^{AB}X_{AB}^{\ell m}$
	and orthogonal in the same sense as the vector harmonics are. The
	separation of variables takes the form \begin{subequations} 
		\begin{align}
			h_{uu} & =\sum_{\ell,m}f_{\ell m}\left(u,r\right)Y_{\ell m},\\
			h_{ur} & =\sum_{\ell,m}h_{\ell m}\left(u,r\right)Y_{\ell m},\\
			h_{uA} & =\sum_{\ell,m}F_{1}^{\ell m}\left(u,r\right)Y_{A}^{\ell m}+G_{1}^{\ell m}\left(u,r\right)X_{A}^{\ell m},\\
			h_{AB} & =\sum_{\ell,m}F_{2}^{\ell m}\left(u,r\right)Y_{AB}^{\ell m}+G_{2}^{\ell m}\left(u,r\right)X_{AB}^{\ell m},
		\end{align}
	\end{subequations} where the sum here is restricted to $\ell\ge2$ and $m$ ranges from $-\ell$ to $\ell$. We neglect the $\ell=0$ and $\ell=1$ multipoles, which are non-radiative and require a special
	treatment. We have also set $h_{rr}=h_{rA}=\gamma^{AB}h_{AB}=0$ to ensure that the linearized metric satisfies the required fall-off in Eq.\eqref{eq:fall-off-linearized}. This gauge choice can always be made (in fact, there is some residual gauge freedom left that we will use in the analysis below). 
	
	The even and odd-parity modes remain decoupled in the linearized Einstein's equation and so are all the $\ell,m$ modes in the spherical
	decomposition. We will restrict ourselves in this section to the $\ell=2$
	modes; the structure of the solution is very similar for higher $\ell$
	modes and we will briefly comment on the form of the general solution
	at the end. Solving for the simpler, odd-parity sector first,
	we find the following retarded solution: 
	\begin{align}
		G_{1}^{\ell=2}\left(u,r\right) & =\sum_{m=-2}^{2}\left[\frac{1}{2}\dot{C}_{2}^{m}\frac{r^{2}}{l^{4}}-\left(l^{-2}\dot{C}_{1}^{m}-\dddot{C}_{1}^{m}\right)+\frac{2}{r}\ddot{C}_{1}^{m}\right.\nonumber \\
		& \;\;\left.+\frac{3}{2r^{2}}\dot{C}_{1}^{m}\right],\\
		G_{2}^{\ell=2}\left(u,r\right) & =\sum_{m=-2}^{2}\left[\left(C_{2}^{m}+C_{1}^{m}-\ddot{C}_{1}^{m}l^{2}\right)\frac{r^{2}}{l^{4}}\right.\nonumber \\
		& \left.+r\left(l^{-2}\dot{C}_{1}^{m}-\dddot{C}_{1}^{m}\right)+\frac{1}{r}\dot{C}_{1}^{m}\right].
	\end{align}
	with $C_{1}^{m},C_{2}^{m}$ dimensions of length squared. The term proportional
	to $r^{2}$ in $G_{2}^{\ell=2,m}$ spoils the fall-off behavior of
	the angular part of the metric in Eq.~\eqref{eq:fall-off-physical}.
	This is, however, easily remedied by realizing that the solution is
	not completely gauge fixed and with the residual gauge freedom this
	part of the metric can be gauged away. With an appropriate gauge choice,
	we set\footnote{By the Stewart-Walker lemma, the linearized Weyl tensor is gauge-invariant
		and a straightforward computation shows that the linearized Weyl tensor
		is independent of $C_{2}^{m}$. Therefore, the $C_{2}^{m}$ solution
		is pure gauge and contains no physical degrees of freedom. This interpretation
		of the solution is consistent with the gauge choice made in Eq.~\eqref{eq:gauge-fix-magnetic}.} 
	\begin{equation}
		C_{2}^{m}\equiv\ddot{C}_{1}^{m}l^{2}-C_{1}^{m}.\label{eq:gauge-fix-magnetic}
	\end{equation}
	This gauge choice is further preserved by the residual gauge freedom generated
	by $\chi^{A}=\epsilon^{AB}D_{B}\left(\vec{\Omega}\cdot\hat{r}\right)$.
	With this gauge choice, and introducing $B_{m}=\dot{C}_{1}^{m}$,
	we finally obtain the following odd-parity solutions for the quadrupolar
	modes with~$\ell=2$ 
	\begin{subequations} \label{eq:magsol} 
		\begin{align}
			h_{uu}^{\text{odd}} & =0\\
			h_{ur}^{\text{odd}} & =0\\
			h_{uA}^{\text{odd}} & =\sum_{m=-2}^{2}\left[\frac{1}{2}\left(\ddot{B}_{m}-l^{-2}B_{m}\right)\frac{r^{2}}{l^{2}}+\left(\ddot{B}_{m}-l^{-2}B_{m}\right)\right.\nonumber \\
			& \left.+\frac{2}{r}\dot{B}_{m}+\frac{3}{2r^{2}}B_{m}\right]X_{A}^{2m}\\
			h_{AB}^{\text{odd}} & =\sum_{m=-2}^{2}\left[r\left(\ddot{B}_{m}-l^{-2}B_{m}\right)-\frac{1}{r}B_{m}\right]X_{AB}^{2m}
		\end{align}
	\end{subequations} where $B_{m}=B_{m}\left(u\right)$ is an arbitrary function of the retarded time $u$ with dimensions of length. Note
	that the leading order of the angular metric is independent of the wave, but that $C_{AB}\neq0$ and $U_{(0)}^{A}\neq0$ (with the constraint
	relating these metric coefficients in Eq.~\eqref{eq:constraint-einstein}
	satisfied).
	
	The solution for general $\ell$ is more complicated, but has these general features: 
	\begin{align}
		G_{1}^{\ell}(u,r) & =\sum_{m=-2}^{2}\left[\frac{1}{2}\dot{C}_{2}^{\ell m}\frac{r^{2}}{l^{4}}+\sum_{i=0}^{\ell}a_{i}^{(\ell)}(r)\;\overset{(i)}{C_{i}^{\ell m}}\right]\\
		G_{2}^{\ell}(u,r) & =\sum_{m=-2}^{2}\left[C_{2}^{\ell m}\frac{r^{2}}{l^{4}}+\sum_{i=0}^{\ell}\left\{ \begin{array}{ll}
			b_{i}^{(\ell)}(r)\;\overset{(i)}{C_{i}^{\ell m}} & \text{if }\ell\text{ even}\\
			b_{i}^{(\ell)}(r)\;\overset{(i+1)}{C_{i}^{\ell m}} & \text{if }\ell\text{ odd}
		\end{array}\right.\right]
	\end{align}
	with the $C$-coefficients depending on $u$ only, the factor $(i)$
	on top of these coefficients indicate its $i$-th derivative with
	respect to $u$ and $a_{i}^{(\ell)},b_{i}^{(\ell)}$ are polynomials
	in $r$ (and its inverse powers) with the highest power being $r^{2}$.
	Note that the term proportional to $C_{2}$ is in fact independent of $\ell$ and can always be gauged away. As a result, even though generically $G_{2}$ contains terms proportional
	to $r^{2}$ which could spoil the desired fall-off, these terms can
	always be set to zero by a clever gauge choice for $C_{2}$ --- similar
	to the case with $\ell=2$. Hence, linearized solutions with odd-parity satisfy the desired fall-off conditions for any $\ell \ge 2$.
	
	The analysis for the even-parity sector mimicks that of the odd-parity sector, but is more involved as more terms are non-zero. Nonetheless,
	also in this case one can gauge fix the solution to obtain
	a linearized solution that satisfies the fall-off conditions prescribed
	in Eq.~\eqref{eq:fall-off-linearized}. Specifically, the retarded
	$\ell=2$ even-parity solutions for $h_{\mu\nu}$ takes the form 
	\begin{subequations}
		\label{eq:elecsol} 
		\begin{align}
			h_{uu}^{\text{even}} & =\sum_{m=-2}^{2}\left[3\left(\ddot{A}_{m}-4l^{-2}A_{m}\right)\frac{r}{l^{2}}\right.\nonumber \\
			& \left.+6\left(\ddot{A}_{m}-l^{-2}A_{m}\right)\frac{1}{r}+\frac{6}{r^{2}}\dot{A}_{m}+\frac{3}{r^{3}}A_{m}\right]Y_{2m},\\
			h_{ur}^{\text{even}} & =0,\\
			h_{uA}^{\text{even}} & =\sum_{m=-2}^{2}\left[\left(2l^{-2}A_{m}-\frac{1}{2}\ddot{A}_{m}\right)\frac{r^{2}}{l^{2}}+\left(4l^{-2}A_{m}-\ddot{A}_{m}\right)\right.\nonumber \\
			& \left.+\frac{2}{r}\dot{A}_{m}+\frac{3}{2r^{2}}A_{m}\right]Y_{A}^{2m},\\
			h_{AB}^{\text{even}} & =\sum_{m=-2}^{2}\left[r\left(\ddot{A}_{m}-4l^{-2}A_{m}\right)+\frac{1}{r}A_{m}\right]Y_{AB}^{2m}
		\end{align}
	\end{subequations} 
	where $A_{m}=A_{m}(u)$ is an arbitrary function of the retarded time $u$ and dimensions of length. Also, similar to the odd-parity sector, we have set $h_{rr}=h_{rA}=\gamma^{AB}h_{AB}=0$. Note that there is backreaction on the ``background'' metric as $r\to \infty$ through the leading term of $h_{uA}$, that is, $U_{(0)}^{A}\neq0$. The backreaction onto the leading order part is unique to $\Lambda\neq0$. In the limit $l\to\infty$, this backreaction vanishes. This is immediately clear from the limit
	of the even- and odd-parity solutions: 
	\begin{subequations} \label{eq:limit-lin-sol}
		\begin{align}
			\lim_{l\to\infty}h_{uu} & =\sum_{m=-2}^{2}\left[\frac{6\ddot{A}_{m}}{r}+\frac{6\dot{A}_{m}}{r^{2}}+\frac{3A_{m}}{r^{3}}\right]Y_{2m},\\
			\lim_{l\to\infty}h_{ur} & =0,\\
			\lim_{l\to\infty}h_{uA} & =\sum_{m=-2}^{2}\left(\left[-\ddot{A}_{m}+\frac{2\dot{A}_{m}}{r}+\frac{3A_{m}}{2r^{2}}\right]Y_{A}^{2m}\right.\nonumber \\
			& \left.+\left[\ddot{B}_{m}+\frac{2\dot{B}_{m}}{r}+\frac{3B_{m}}{2r^{2}}\right]X_{A}^{2m}\right),\\
			\lim_{l\to\infty}h_{AB} & =\sum_{m=-2}^{2}\left(\left[\frac{\ddot{A}_{m}}{r}+\frac{A_{m}}{r^{3}}\right]r^{2}Y_{AB}^{2m}\right.\nonumber \\
			& \left.+\left[\frac{\ddot{B}_{m}}{r}-\frac{B_{m}}{r^{3}}\right]r^{2}X_{AB}^{2m}\right)\;,
		\end{align}
	\end{subequations} where $A_{m}$ and $B_{m}$ reduce to the standard quadrupole moments on flat spacetime.
	
	Connecting these results with the Bondi-Sachs expansions, we find
	that the linear part of the metric coefficients
	is given by 
	\begin{subequations} 
		\begin{align}
			M & =\sum_{m=-2}^{2}3\left(\ddot{A}_{m}-l^{-2}A_{m}\right)Y^{2m}\\
			U_{A}^{(0)} & =\frac{1}{l^{2}}\sum_{m=-2}^{2}\left(2l^{-2}A_{m}-\frac{1}{2}\ddot{A}_{m}\right)Y_{A}^{2m}\nonumber \\
			& +\frac{1}{2}\left(\ddot{B}_{m}-l^{-2}B_{m}\right)X_{A}^{2m}\\
			N_{A} & =\sum_{m=-2}^{2}-3\dot{A}_{m}\;Y_{A}^{2m}+3\dot{B}_{m}X_{A}^{2m}\\
			C_{AB} & =\sum_{m=-2}^{2}\left(\ddot{A}_{m}-4l^{-2}A_{m}\right)Y_{AB}^{2m}+\left(\ddot{B}_{m}-l^{-2}B_{m}\right)X_{AB}^{2m}\\
			E_{AB} & =\sum_{m=-2}^{2}A_{m}\;Y_{AB}^{2m}-B_{m}\;X_{AB}^{2m}
		\end{align}
	\end{subequations} and all other coefficients vanishing or determined
	by lower order terms. \\
	\\
	The radiation rate at the linearized level in Eq.~\eqref{eq:Energy-linearized}
	reduces after some further simplifications to 
	\begin{align}
		\frac{dE}{du} & =-\frac{3}{8\pi G}\sum_{m=-2}^{2}\left[\left|\dddot{A}_{m}-4l^{-2}\dot{A}_{m}\right|^{2}\right.\nonumber \\
		& \left. -3l^{-2}\ddot{A}_{m}^*\left(\ddot{A}_{m}-4l^{-2}A_{m}\right)+\left|\dddot{B}_{m}-l^{-2}\dot{B}_{m}\right|^{2} \right. \nonumber \\
		& \left. +3l^{-2}\ddot{B}_{m}^*\left(\ddot{B}_{m}-l^{-2}B_{m}\right) \right]\;, \label{eq:flux-energy-lin}
	\end{align}
	where the star indicates complex conjugation.
	If we consider the total energy
	radiated during some large time interval $\Delta u$, where we assume
	that at far past and at far future the system will not radiate so
	that we can remove the boundary terms in time, then 
	\begin{align}
		E_{\Delta u} & =-\frac{3}{8\pi G}\int_{-\infty}^{\infty}du\sum_{m=-2}^{2}\left[\left(\dddot{A}_{m}-\frac{\dot{A}_{m}}{l^{2}}\right)\right.  \\
		& \left.\left(\dddot{A}_{m}^*-\frac{4\dot{A}_{m}^*}{l^{2}}\right)+\left(\dddot{B}_{m}-\frac{\dot{B}_{m}}{l^{2}}\right)\left(\dddot{B}^*_{m}-\frac{4\dot{B}^*_{m}}{l^{2}}\right)\right]. \nonumber
	\end{align}
	In particular, after integration by parts, 
	we have 
	\begin{align}
		E_{\Delta u} & =-\frac{3}{8\pi G}\int_{-\infty}^{\infty}du\sum_{m=-2}^{2}\left[\left|\dddot{A}_m\right|^{2}+\frac{5|\ddot{A}_m|^{2}}{l^{2}}+\frac{4|\dot{A}_m|^{2}}{l^{4}}\right.\nonumber \\
		& \left.+\left|\dddot{B}_m\right|^{2}+\frac{5|\ddot{B}_m|^{2}}{l^{2}}+\frac{4|\dot{B}_m|^{2}}{l^{4}}\right] . \label{eq:energy-flux-quad}
	\end{align}
	This flux is manifestly negative. Therefore, the total energy always decreases
	for a source characterized by a quadrupole. The flat spacetime limit
	yields the expected result for a quadrupolar source $A,B$ 
	\begin{equation}
		\lim_{l\to\infty}E_{\Delta u}=-\frac{3}{8\pi G}\int_{-\infty}^{\infty}du \sum_{m=-2}^{2}\left[\left|\dddot{A}_{m}\right|^{2}+\left|\dddot{B}_{m}\right|^{2}\right].
	\end{equation}
	Note that for $\Lambda<0$, i.e. $l \to i l$, the energy flux is non-zero so that the boundary is not reflective (as is typically imposed). In fact, the energy flux can have an arbitrary sign depending on the values of $A$, its time derivatives and $l$. 

	\subsection{Robinson-Trautman spacetime}
	\label{sec:RT-solution}
	
	The Robinson-Trautman spacetime is an exact solution of Einstein equations that describes the backreaction of a non-linear gravitational wave on a Schwarzschild spacetime. 
	The Robinson-Trautman solution is dynamical: it models gravitational
	radiation expanding from a radiating object. Since ultimately, we
	are interested in describing gravitational radiation emitted by compact
	sources in the presence of a cosmological constant, this example is
	of particular interest for our analysis. The original Robinson-Trautman solution
	contained no cosmological constant \citep{Robinson:1962zz}, but it
	was soon realized that the solution easily accommodates for a non-zero
	cosmological constant. This class of spacetimes is the most
	general radiative vacuum solution admitting a geodesic, shear-free
	and twist-free null congruence of diverging rays. It has been shown
	that starting with arbitrary, smooth initial data at some retarded
	time $u=u_{0}$ 
	, the cosmological
	Robinson-Trautman solutions converge exponentially fast to a Schwarzschild-de
	Sitter solution at large retarded times ($u\to\infty$). Thus, these
	solutions also belong to the class of solutions discussed in this
	paper. In this section, we will show this explicitly by providing
	the form of this solution in Bondi-Sachs like coordinates.
	
	The line element of the Robinson-Trautman solution with a positive
	cosmological constant is given by
	
	\begin{align}
		ds^{2}=&-2H\left(u,r,\theta,\phi\right)du^{2}-2dudr\nonumber\\
		&+\frac{r^{2}}{P^{2}\left(u,\theta,\phi\right)}\left(d\theta^{2}+\sin^{2}\theta d\phi^{2}\right),\label{eq:RT-solution}  
	\end{align}
	with 
	\[
	2H\left(u,r,\theta,\phi\right)=-\frac{r^{2}}{\ell^{2}}-\frac{2r\dot{P}}{P}+\frac{1}{2}\mathcal{R}_{h}-\frac{2m\left(u\right)}{r}.
	\]
	Here $P\left(u,\theta,\phi\right)$ is an arbitrary function of the
	retarded time and the angles, and contains the information of the gravitational wave. According to Einstein's equations, the following equation governs the time evolution of $m\left(u\right)$:
	\begin{equation}
		\dot{m}=3m\frac{\dot{P}}{P}-\frac{1}{8}\Delta_{h}\mathcal{R}_{h}.\label{eq:EOM-2}
	\end{equation}
	The Laplacian $\Delta_{h}$ is defined with respect to the metric $h_{AB}=P^{-2}\left(u,\theta,\phi\right)\left(d\theta^{2}+\sin^{2}\theta d\phi^{2}\right)$. In the particular case when $P=1$ (no radiation), the Schwarzschild
	de Sitter solution is recovered. 
	
	The Robinson-Trautman solution, as written in Eq.~\eqref{eq:RT-solution},
	does not fit immediately within the asymptotic conditions in Eqs.~\eqref{eq:fall-off-physical}. The reason is
	the presence of the function $P^{-2}$ appearing in front of the metric of the
	2-sphere. In order to accommodate the solution one must perform an appropriate change of coordinates. In general, the implementation of this change of coordinates is technically a very hard task. However,
	a simplified analysis can be achieved by considering
	the Robinson-Trautman metric with axial symmetry. In addition, and
	for clarity to the reader, in this section we will only consider a linearized
	version of the solution. The non-linear analysis will be discussed in App.~\ref{app:RT-coordinate-change}.

	
	Assuming for simplicity axial symmetry, the linearized Robinson-Trautman solution expanded around a Schwarzschild-de Sitter background is obtained by expressing the function $P=P\left(u,\theta\right)$ as follows:
	\[
	P=1+\epsilon p\left(u,\theta\right).
	\]
	Here $\epsilon$ is a small parameter that controls the linearized
	expansion. In this approximation, the leading order of Eq.~\eqref{eq:EOM-2} becomes  
	\begin{align}
		\dot{m}=&\frac{1}{4}\epsilon \left[12m\dot{p}+p^{(4)}-\cot^{2}\theta p''+2\cot\theta p'''\right.\nonumber\\
		&\left.+\cot\theta\left(\csc^{2}\theta+2\right)p'\right]+O\left(\epsilon^{2}\right).
	\end{align}
	Here the prime denotes derivatives with respect to $\theta$.
	In order to accommodate the solution within our asymptotic conditions, one can implement the following change of coordinates to linear
	order in $\epsilon$ 
	\begin{align*}
		u & \rightarrow u-\epsilon f\left(\mathit{u},\theta\right)+O\left(r^{-3}\right),\\
		r & \rightarrow r\left(1+\epsilon p\left(\mathit{u},\theta\right)\right)-\frac{1}{2}\epsilon D^{2} f\left(\mathit{u},\theta\right)+O\left(r^{-2}\right),\\
		\theta & \rightarrow\theta+\epsilon\frac{f'\left(\mathit{u},\theta\right)}{r}+O\left(r^{-3}\right),\\
		\phi & \rightarrow\phi,
	\end{align*}
	where
	\begin{equation}
		p=\dot{f}.\label{eq:condlin}
	\end{equation}
	Thus, one finds
	\begin{subequations} \label{eq:RTcoeff}
		\begin{align}
			&M  =m\left[1-\epsilon\left(3\dot{f}+f\dot{m}\right)\right],\\
			&C_{\theta\theta} =\epsilon\left(f''-\cot\theta f'\right), \; \; C_{\phi\phi}=-\epsilon\sin^{2}\theta\left(f''-\cot\theta f'\right),\\ 
			& C_{\theta\phi}=0,\\
			&U_{\left(0\right)}^{\theta}=  \frac{\epsilon}{\ell^{2}}f',\;\;\;\; U_{\left(0\right)}^{\phi}=0,\\
			&N^{\theta}=  -3\epsilon mf',\;\;\;\; N^{\phi}=0,\\
			&E_{AB}  =0.
		\end{align}
	\end{subequations} 
	Decomposing $f$ into spherical harmonics $f=\sum_l f_l Y_{l0}$, we can also compactly write  $C_{AB}$ as $C_{AB} = 2 \epsilon \sum_{l} f_l Y^{l0}_{AB}$.
	In addition, at linear order the News tensor in Eq.~\eqref{eq:News} takes the simple form $N_{AB}^{(\Lambda)}=\dot{C}_{AB}$.
	
	To obtain the flux of energy one can replace~\eqref{eq:RTcoeff} in Eq.~\eqref{eq:radrate}, while retaining only the terms up to order $\epsilon^{2}$.
	After some integrations by parts one finally obtains
	\begin{equation}
		\frac{dE}{du}=-\frac{\epsilon^{2}}{16\pi G}\oint d^{2}S\left[\left(\dot{f}''-\cot\theta\dot{f}'\right)^{2}+\frac{3m}{\ell^{2}}\partial_{u}\left(f'^{2}\right)\right]. \label{eq:flux-RT-lin}
	\end{equation}


	\section{Comparison with alternative approaches}
	\label{sec:comparison} 
	
	Given the observational evidence for an accelerated expansion of our
	Universe and the recent gravitational wave observations, the challenge
	of understanding gravitational waves in the presence of a positive
	cosmological constant $\Lambda$ has received considerable attention
	in recent years. At the linearized level, most of the previous results in the literature agree with each other. As we will see below, our results are also in agreement with them.
	
	However, the situation is drastically different in the full non-linear theory.
	Different methods and/or different boundary conditions are employed --- some of which even require regularization; the results in general do not agree. We describe below some of these approaches without any pretense of being exhaustive. 
	
	\subsection{Linearized gravity}
	
	An important starting point is a thorough understanding of weak gravitational
	waves on a de Sitter background. There are two key issues predominantly
	studied within this context: (1) a mathematically sound and physically
	sensible notion of energy and its flux, and (2) finding explicit solutions
	for gravitational waves generated by a compact source and their link
	to time-changing quadrupole moments, thereby generalizing the well-known
	flat result.\footnote{The gravitational memory effect in de Sitter spacetimes has been investigated
		in \citep{Bieri:2015jwa}, however, this paper focused on the cosmological
		horizon and is therefore more difficult to relate to the results in
		this paper that exclusively apply near $\scri$.}
	
	There are various notions of energy and its flux in the literature
	that mostly distinguish themselves by the method used to derive it
	(as a consequence, these notions typically are equivalent up to boundary
	terms), ``where'' in spacetime the energy (flux) is evaluated (mostly
	on $\scri$ or across the cosmological horizon) and by the class of
	linearized solutions for which the energy is defined. For instance,
	in \citep{Kolanowski:2020wfg}, the energy flux across $\scri$ is
	derived using the same symplectic methods as in the earlier work
	in \citep{abk2}, but for a slightly larger class of linearized solutions. In \citep{Kolanowski:2021hwo}, the authors use the Wald-Zoupas prescription to define energy (and angular momentum).  
	In all these cases, the resulting energy flux is finite and the result
	gauge invariant. 

	On the other hand, the energy flux obtained in \citep{Chrusciel:2020rlz}  by direct use of
	Noether currents is not finite (note also earlier work
	\citep{Chrusciel:2016oux}). In that paper, they remedy this issue by isolating the terms which would lead to infinite energy and introducing a ``renormalized canonical energy''. Their argument for the plausibility
	of this procedure is based on the observation that the diverging terms
	have dynamics of their own, which evolves independently from the remaining part of the canonical energy. 
	
	Yet another approach, which
	applies only for sources supporting the short wavelength approximation,
	as it relies on the Isaacson effective stress-tensor, matches the
	results in \citep{abk3} if one identifies the transverse-traceless
	gauge with a certain projection operation.\footnote{For linearized solutions on Minkowski spacetime, this is a well-defined and consistent procedure for the leading order components of the gravitational
		field. This is shown explicitly in \citep{ab}, in which the first
		notion is referred to as `TT' gauge and the second as `tt' gauge.
		However, it is not clear that the two notions are also equivalent
		for the leading order fields in de Sitter spacetime.} 
	
	In \citep{Hoque:2018byx}, the authors employ the
	symplectic current density of the covariant phase space
	to show that the integrand in the energy flux expression on the cosmological horizon is same as that on $\scri$ . This result is interesting as it suggests that at the linearized level propagation
	of energy flux is along null rays in de Sitter spacetime, despite the fact that
	gravitational waves themselves have tail terms due to back-scattering
	off the background curvature. 
	
	The second key issue investigated is the link between time variation
	of some compact source generating gravitational waves and the resulting
	gravitational waves themselves. This was investigated in \citep{abk3}
	by solving the linearized Einstein's equation on de
	Sitter background sourced by a (first order) stress-energy tensor.
	To study the limit to $\scri$, the authors introduce a late time approximation in addition to the commonly used post-Newtonian approximation.
	This allowed them to express the leading terms of the gravitational waves in terms of the quadrupole moments of sources. Moreover, the
	energy carried away by this gravitational waves was studied using
	Hamiltonian methods on the covariant phase-space of the linearized
	solutions introduced in their earlier paper \citep{abk2}. This showed
	that despite the fact that in principle the energy for linearized
	perturbations on de Sitter spacetime can be negative (note that this
	is not in contrast with the finiteness discussed in the previous paragraph),
	the energy of gravitational waves emitted by compact objects is always
	positive. This is also consistent with our results in Sec.~\ref{sec:example-linear-waves}.
	The quadrupolar solutions in \citep{abk3} were also reinterpreted
	in \citep{He:2018ikd}, by writing the solutions in Bondi-Sachs type
	coordinates different from the ones introduced in this paper. The
	authors showed that the quadrupolar solutions can be accommodated
	by a non-zero shear for the leading order part of $g_{AB}$. This
	is different but not in contradiction with the results in this paper,
	which show that the radiative solution contributes to the sub-leading
	part of $g_{AB}$ \emph{and} to $U^A_{(0)}$, while the leading order part of $g_{AB}$ is equal to  $\gamma_{AB}$. This is a gauge choice. Other papers relating the source dynamics modeled by some compact stress-energy tensor to the gravitational wave and the energy have relied on the short
	wave approximation \citep{Date:2016uzr,Hoque:2017xop,Hoque:2018dcg}.
	These results are consistent with the results in \citep{abk2}.
	
	The gravitational memory effect, which describes the permanent displacement of test masses after a gravitational wave has passed, has only sparsely been analyzed in de Sitter spacetimes. This interesting physical effect was studied near the cosmological horizon of de Sitter spacetime in \cite{Bieri:2015jwa,Hamada:2017gdg}, and a ``linear'' memory effect has been linked to the tail of the de Sitter Green's function in \cite{Chu:2016qxp}. However, the connection to asymptotic symmetries --- which exist for the memory effect in asymptotically flat spacetimes --- have not been studied. 
	We intend to explore this in upcoming research.

	\subsection{Full non-linear theory}
	
	Early investigations of the asymptotic structure of asymptotically
	de Sitter spacetimes in full non-linear general relativity such as
	\citep{Strominger:2001pn,Anninos:2010zf,Anninos:2011jp} imposed too
	stringent boundary conditions by demanding conformal flatness of the
	induced three-dimensional metric on $\scri$. As a result, these early
	investigations concluded that the asymptotic symmetry group is the full de Sitter group. However, as was shown in \citep{abk1}, imposing
	conformal flatness ruled out many physically relevant spacetimes as
	they enforced a vanishing flux of radiation across $\scri$ (and consequently all charges are strictly conserved, see also \citep{aneesh_conserved_2019}).

	The observation that demanding the asymptotic symmetry group to be the de Sitter one ruled out gravitational waves sparked new interest in this challenging problem.
	It lead the authors in \citep{He:2015wfa} to consider Bondi-Sachs
	type coordinates for asymptotically de Sitter spacetimes, which are
	not conformally flat at $\scri$. A nice property of their coordinates
	is that the Weyl tensor has peeling behavior near $\scri$ \citep{Xie:2017uqa}.
	While these authors also rely on the Bondi framework, their fall-off
	conditions on the metric coefficients are different from those considered
	here. In particular, the authors did not fix $g_{AB}$ to be equal
	to the unit two-sphere but instead allowed for a non-zero shear at
	leading order. Their shear contains all the information about gravitational
	radiation. However, based on our analysis, using their fall-off conditions
	the variation of the charge is infinite.
	This makes these fall-off conditions
	not as attractive. Moreover, the analysis in those papers was restricted to axi-symmetric spacetimes and limited to the study of Einstein's equations; these papers did not study gravitational charges and fluxes.
	
	Subsequently, various authors used the Newman-Penrose formalism to
	define and study asymptotically de Sitter spacetimes \citep{Saw:2016isu,Saw:2017hsf,Mao:2019ahc}.
	The two earlier papers by Saw used a special choice of null foliation,
	thereby excluding the Robinson-Trautman spacetime with a positive
	cosmological constant as part of their allowed class of spacetimes. The class of null foliations was generalized
	in \citep{Mao:2019ahc} by Mao to accommodate for the Robinson-Trautman
	spacetime. A nice feature of the fall-off conditions on the spin coefficients
	and Weyl scalars in those papers is that they have a well-defined flat limit. 
	However, Mao finds that the asymptotic symmetry algebra consists of all diffeomorphisms on the two-sphere and translations in the $u$-direction.
	In the limit $l\to\infty$, the asymptotic symmetry algebra becomes the algebra of all diffeomorphisms on the two-sphere and supertranslations
	known as the extended BMS algebra \citep{Campiglia:2020qvc} instead
	of the BMS algebra. Gravitational charges and fluxes were not studied in these papers.
	
	Another set of recent papers on this topic uses similar techniques to those used here \citep{Compere:algebra,Compere:group}. These authors find that the asymptotic symmetries form a Lie algebroid
	instead of a Lie algebra, as they used different fall-off conditions on the metric. Their asymptotic
	symmetries consist of infinite-dimensional ``non-abelian supertranslations''
	and superrotations, and like in \citep{Mao:2019ahc} reduces in the
	limit $l\to\infty$ to the extended BMS algebra. These boundary conditions were used in \citep{erfani_bondi_2022} to define a Bondi news-like tensor using a Newman-Penrose tetrad.
	
	Inspired by the dictionary between Bondi and Fefferman-Graham gauges \citep{Poole:2018koa},
	the authors used earlier results in Fefferman-Graham gauges to define a new class of asymptotically de Sitter spacetimes. 
	In their follow-up work \citep{poole_charges_2022}, in order to obtain finite charges and fluxes, these authors introduce a holographic renormalization procedure while all charges and fluxes
	are naturally finite in this paper and do not require any \emph{ad
		hoc} regularization. The latter work also states more clearly that their interest is in spacetimes with compact spatial slices, as opposed to this work.
	
	Other work has focused on studying the possible isometries of asymptotically de Sitter spacetimes. One of the key results is that the asymptotic symmetry algebra they find is maximally four-dimensional \cite{Kaminski:2022tum}, which in spirit agrees with our work. 
	
	Research in a different direction focused on the question of how to
	identify the presence of gravitational radiation in the presence of
	$\Lambda$ using geometric tools only and without referring to a specific
	coordinate system \citep{radiation-criterion}. The criterion proposed
	is based on value of super-Poynting vector at $\scri$: if it vanishes,
	there is no gravitational radiation across $\scri$ while if it is
	non-zero, there is gravitational radiation across $\scri$. This criterion
	is straightforward to check as the super-Poynting vector is the commutator
	of the leading order electric $\mathcal{E}_{ab}$ and magnetic $\mathcal{B}_{ab}$
	part of the Weyl tensor. When the cosmological constant vanishes,
	this criterion is equivalent to the standard `identification' method
	of gravitational radiation at null infinity through the means of the
	(non-)vanishing of the Bondi news tensor \citep{radiation-criterion-flat}.
	When $\mathcal{B}_{ab}$ vanishes on $\scri$, the super-Poynting
	vector also vanishes and this criterion implies that there is no radiation.
	However, the vanishing of $\mathcal{B}_{ab}$ is not a necessary condition.
	In particular, certain Kerr-de Sitter generalized spacetimes have
	non-vanishing $\mathcal{B}_{ab}$ on $\scri$ yet their super-Poynting
	vector vanishes. Here we find that there is gravitational radiation
	whenever $C_{AB}$ (and hence $U_{(0)}^{A}$) is non-zero. When these
	are zero, $\mathcal{B}_{ab}$ vanishes. Therefore, our criterion to
	establish the presence of gravitational radiation seems to be stricter.
	In other words, based on the super-Poynting vector criterion a spacetime
	may be labeled as non-radiating, while based on $C_{AB}$ it would
	be considered radiating.
	It is therefore not too surprising that the authors in follow-up work found that the asymptotic symmetry algebra for the spacetimes they considered are infinite-dimensional \citep{Fernandez-Alvarez:2021yog,Fernandez-Alvarez:2021zmp,Senovilla:2022pym}.

	\begin{acknowledgments}
		BB would like to thank Ahbay Ashtekar for many thought-provoking conversations over the years, as well as CECs for its hospitality during her visits, separated in time by force majeure, when this work was initiated and completed. 
		CB would like to thank Neil Turok for bringing the authors of this paper together at the Path Integral for Gravity workshop held at the Perimeter Institute in November 2017, when the discussions that led to this article started.
		AP wishes to thank Jorge Nore\~na and Ricardo Troncoso for helpful discussions. 
		We also would like to thank Amitabh Virmani for pointing out Ref.~\cite{Teukolsky:1982nz} as well as a few typos in the preprint version of this paper, and Jos\'e Senovilla, Francisco \'Alvarez and Kartik Prabhu for a productive email exchange, which lead to improvements of our manuscript.
		The research of AP is partially supported by Fondecyt grants No 1211226, 1220910 and 1230853.
	\end{acknowledgments}

	\appendix
	
	\section{Link with geometric approach}
	\label{app:geometry}
	The goal of this appendix is to show how the metric in Eq.~\eqref{eq:metric-Bondi-gauge} with the fall-off conditions specified in Eq.~\eqref{eq:fall-off-physical} is related to earlier results obtained using purely geometric techniques in \citep{abk1,abk2,abk3,abk:prl}. In particular, here we show that the fall-off conditions are such that the class of spacetimes described here are not \emph{strongly} asymptotically de Sitter spacetimes. This is as desired, because strongly asymptotically de Sitter spacetimes have no radiative fluxes across $\scri$.
	\\
	\\
	From the coordinates and expression for the physical metric in the main body of this paper, we can construct Bondi-Sachs type coordinates
	near $\scri$ for the conformally completed spacetime $\tilde{g}_{\mu \nu} = \Omega^2 g_{\mu \nu}$ with $r=\Omega^{-1}$. 
	In particular, we straightforwardly obtain
	\begin{align} \label{eq:metric-at-scri}
		\tilde{g}_{\mu\nu}dx^{\mu}dx^{\nu} \hateq & \; UWdu^{2}  \\
		&+\gamma_{AB}\left(dx^{A}-U^{A}du\right)\left(dx^{B}-U^{B}du\right),\nonumber
	\end{align}
	where the fall-off of the coefficients directly follows from that in Eq.~\eqref{eq:fall-off-physical} 
	\begin{subequations}\label{eq:fall-off-conformal} 
		\begin{align}
			\beta & =-\frac{1}{32} C^{AB}C_{AB} \Omega^{2} \nonumber \\
			& \;\;+\frac{1}{128}\left(\left(C^{AB}C_{AB}\right)^{2}-12C^{AB}E_{AB}\right) \Omega^4+\ldots, \\
			W & =\frac{\Lambda}{3}-D_{A}U_{\left(0\right)}^{A}\; \Omega -\left(1+\frac{\Lambda}{16}C^{AB}C_{AB}\right) \Omega^2 \nonumber \\
			& \qquad +2M \Omega^3 + \ldots\\
			U^{A} & =U_{(0)}^{A}-\frac{1}{2}D_{B}C^{AB} \Omega^2 \nonumber \\
			& \qquad -\frac{2}{3} \left(N^{A}-\frac{1}{2}C^{AB}D^{C}C_{BC}\right) \Omega^3 +\ldots, \\
			q_{AB} & =\gamma_{AB}+C_{AB}\Omega+\frac{C^{CD}C_{CD}\gamma_{AB}}{4r^{2}} \Omega^{2}+E_{AB}\Omega^{3}+\ldots .
		\end{align}
	\end{subequations} 
	The surfaces of constant $u$ are outgoing null surfaces for
	both the unphysical and the physical spacetime, since conformal transformations	do not change the properties of null geodesics. 
	With this choice of the fall-off conditions and conformal factor, the
	divergence of $\tilde{\grad}_{\mu}\Omega$ is non-zero. This
	is different from the typical choice made in the asymptotically flat
	context, where one often chooses $\Omega$ such that one is in a conformal
	divergence-free frame because for asymptotically flat spacetimes this choice simplifies intermediate results significantly.	

	In \citep{abk1}, it was shown that strongly asymptotically de Sitter spacetimes have a conformally flat metric at $\scri$ and all fluxes vanish across $\scri$. It is clear from Eq.~\eqref{eq:metric-at-scri} that the class of spacetimes considered in this paper do not belong to strongly asymptotically de Sitter spacetimes. This can also been seen by studying the Weyl tensor. In particular, conformal flatness of the metric at $\scri$ is equivalent to the vanishing of the next-to-leading order magnetic part of the Weyl tensor at $\scri$.
	Given the above expression for the metric, we can compute the Weyl tensor explicitly.
	We find that -- as expected -- the leading order part of the Weyl tensor vanishes on $\scri$, that is, $C_{abcd}\hateq0$ \citep{abk1}. The next-to-leading order part is non-zero and we will decompose it into its electric and magnetic part: 
	\begin{align}
		\mathcal{E}_{ab} & :=l^{2}\Omega^{-1}C_{acbd}n^{c}n^{d}\label{eq:defE}\\
		\mathcal{B}_{ab} & :=l^{2}\Omega^{-1}{}^{*}C_{acbd}n^{c}n^{d}=\frac{l^{2}}{2}\epsilon_{ac}^{\;\;\;mn}\;C_{mnbd}n^{c}n^{d}\;.\label{eq:defB}
	\end{align}
	where both $\mathcal{E}_{ab}$ and $\mathcal{B}_{ab}$ are symmetric,
	traceless and orthogonal to $n^{a}$, which here is equal to $n^{a}\partial_{a}=\partial/\partial u$.
	The resulting expressions are rather long, so here we only show the part linear in the metric coefficients:
	\begin{align}
		\left.\mathcal{E}_{ab}dx^{a}dx^{b}\right|_{{\rm lin}} & \hateq-\frac{1}{l^{2}}V^{(3)}\;\left(du+l^{2}d\Omega\right)^{2}\nonumber \\
		& +\left(\frac{3}{2l^{2}}U_{A}^{(3)}+\frac{l^{2}}{2}\partial_{u}\left[D^2+1\right]U_{A}^{(0)}\right)\left(du\right. \nonumber \\
		& +\left.\ell^{2}d\Omega\right)dx^{A}-\frac{l^{2}}{2}\left(\frac{3}{l^{4}}q_{AB}^{(3)}-\frac{1}{l^{2}}V^{(3)}\gamma_{AB}\right. \nonumber \\
		& +\left[2l^{2}\partial_{u}^{2}+D^2-4\right]\left[D_{(A}U_{B)}^{(0)}-\frac{1}{2}\gamma_{AB}D_{C}U_{(0)}^{C}\right]\nonumber \\
		& \left.-\left[D_{(A}D_{B)}-\frac{1}{2} \gamma_{AB} D^2 \right]D_{C}U_{(0)}^{C}\right)dx^{A}dx^{B}\label{eq:lin-electric}
	\end{align}
	and 
	\begin{align}
		\left.\mathcal{B}_{ab}dx^{a}dx^{b}\right|_{{\rm lin}} & \hateq\frac{1}{2}\left[D^2+2\right]\epsilon^{AB}D_{A}U_{B}^{(0)}\left(du+l^{2}d\Omega\right)^{2}\nonumber \\
		& \frac{l^{2}}{2}\partial_{u}\left[D^2+1\right]\epsilon_{A}^{\;\;B}U_{B}^{(0)}(du+l^{2}d\Omega)dx^{A} \nonumber \\
		& -\frac{l^{2}}{2}\left(\frac{1}{2}\left[D^2+2\right](\epsilon^{EF}D_{E}U_{F}^{(0)})\;\gamma_{AB}\right.\nonumber \\
		& -2l^{2}\partial_{u}^{2}\left[D_{(A}(\epsilon_{B)}^{\;\;\;\;C}U_{C}^{(0)})
		\right. \nonumber \\
		& \left.-\frac{1}{2}\gamma_{AB}D_{C}(\epsilon^{CE}U_{E}^{(0)})\right] \nonumber \\
		& \left.-\left[D_{(A}D_{B)}-\frac{1}{2}\gamma_{AB}D^{2}\right]\epsilon^{EF}(D_{E}U_{F}^{(0)})\right)\times \nonumber \\
		& \times dx^{A}dx^{B}\;. \label{eq:lin-magnetic}
	\end{align}
	It is evident that $\mathcal{B}_{ab}$ is non-zero and consequently the class of spacetimes considered in this paper are not strongly asymptotically de Sitter; as desired, because strongly asymptotically spacetimes remove half the permissible data and have no fluxes of energy across $\scri$. 
	
	Using the explicit linearized solutions in Sec.~\ref{sec:example-linear-waves}, we find that for quadrupolar gravitational waves, the electric and magnetic part of the Weyl tensor are 
	\begin{align}
		\mathcal{E}_{ab}dx^{a}dx^{b} & \hateq\sum_{m=-2}^{2}\left\{ 6l^{-2}\left(l^{-2}A_{m}-\ddot{A}_{m}\right)Y_{2m}\;du^{2}\right.\nonumber \\
		& +\left(\del_{u}\left(l^{-2}A_{m}-\ddot{A}_{m}\right)Y_{A}^{2m}\right.\nonumber \\
		& \left.-\del_{u}\left(4l^{-2}B_{m}-\ddot{B}_{m}\right)X_{A}^{2m}\right)dudx^{A}\nonumber \\
		& +\left(\left[-\frac{3}{2}\left(l^{-2}A_{m}-\ddot{A}_{m}\right)\right.\right.\nonumber \\
		& \left.+\frac{1}{2}\del_{u}^{2}\left(l^{-2}A_{m}-\ddot{A}_{m}\right)\right]Y_{AB}^{2m}\nonumber \\
		& -3\left(l^{-2}A_{m}-\ddot{A}_{m}\right)Y_{2m}\;S_{AB}\nonumber \\
		& \left.\left.-\frac{1}{2}\del_{u}^{2}\left(4B_{m}-l^{2}\ddot{B}_{m}\right)X_{AB}^{2m}\right)dx^{A}dx^{B}\right\} \label{eq:lin-electric-explicit}
	\end{align}
	and 
	\begin{align}
		\mathcal{B}_{ab}dx^{a}dx^{b} & \hateq\sum_{m=-2}^{2}\left\{ 6l^{-2}\left(l^{-2}B_{m}-\ddot{B}_{m}\right)Y_{2m}\;du^{2}\right.\nonumber \\
		& -\left(\del_{u}\left(4l^{-2}A_{m}-\ddot{A}_{m}\right)Y_{A}^{2m}\right.\nonumber \\
		& \left.-\del_{u}\left(l^{-2}B_{m}-\ddot{B}_{m}\right)X_{A}^{2m}\right)dudx^{A}\nonumber \\
		& +\left(\left[-\frac{3}{2}\left(l^{-2}B_{m}-\ddot{B}_{m}\right)\right.\right.\nonumber \\
		& \left.+\frac{1}{2}\del_{u}^{2}\left(l^{-2}B_{m}-\ddot{B}_{m}\right)\right]Y_{AB}^{2m}\nonumber \\
		& \left.-3\left(l^{-2}B_{m}-\ddot{B}_{m}\right)Y_{2m}\;S_{AB}\right.\nonumber \\
		& \left.\left.+\frac{1}{2}\del_{u}^{2}\left(4A_{m}-l^{2}\ddot{A}_{m}\right)X_{AB}^{2m}\right)dx^{A}dx^{B}\right\} \label{eq:lin-magnetic-explicit}
	\end{align}
	We have explicitly verified that these expressions are symmetric, transverse $\mathcal{E}_{ab}n^{b}\hateq0\hateq\mathcal{B}_{ab}n^{b}$
	and traceless $q^{ab}\mathcal{E}_{ab}\hateq0\hateq q^{ab}\mathcal{B}_{ab}$. In taking the limit $l\to\infty$, one needs to be
	careful to rescale $\mathcal{E}_{ab}$ and $\mathcal{B}_{ab}$; otherwise,
	due to the overall factor of $l^{2}$ in the definition in Eqs.~\eqref{eq:defE}-\eqref{eq:defB},
	this limit trivially diverges. The flat limit is 
	\begin{align}
		\lim_{l\to\infty}l^{-2}\mathcal{E}_{ab}dx^{a}dx^{b} & \hateq\sum_{m=-2}^{2}-2\del_{u}^{4}B_{m}X_{AB}^{2m}dx^{A}dx^{B}\\
		\lim_{l\to\infty}l^{-2}\mathcal{B}_{ab}dx^{a}dx^{b} & \hateq\sum_{m=-2}^{2}2\del_{u}^{4}A_{m}X_{AB}^{2m}dx^{A}dx^{B}\;.
	\end{align}
	Note that the parity-even solution, which is sometimes also called an electric solution, contributes to the magnetic part of
	the Weyl tensor and vice versa. There is no contradiction here, as
	the names electric and magnetic refer to very different notions.
	
	In this linearized limit one can also explicitly show that the $\ell=1$
	modes in $U_{(0)}^{A}$ do not contribute to $\mathcal{E}_{ab}$ nor
	to $\mathcal{B}_{ab}$.\footnote{To show this, first decompose $U_{A}^{(0)}$ into an `electric' and `magnetic' part: $U_{A}^{(0)}=D_{A}f+\epsilon_{A}^{\;\;B}D_{B}g$ and use that if $f$ and $g$ are $\ell=1$ modes $D_{A}D_{B}f=-\gamma_{AB}f$
		and similarly for $g$.} This further supports the interpretation of those modes as non-radiative.

	\section{Change of coordinates for the non-linear Robinson-Trautman solution}
	\label{app:RT-coordinate-change}	
	In this appendix we show that the non-linear Robinson-Trautman solution can be accommodated within the asymptotic conditions in Eq.~\eqref{eq:fall-off-physical}. Starting from the solution in Eq.~\eqref{eq:RT-solution} one can perform the following asymptotic change of coordinates
	\begin{align*}
		u  \rightarrow & f\left(u,\theta\right)+\frac{\ell^{2}\csc\theta\left(\sin\theta h'-\sin h\right)}{P}\frac{1}{r}+\frac{U_{2}}{r^{2}}+\dots,\\
		r \rightarrow & \sin\theta\csc hP\,r+R_{0}+\frac{R_{1}}{r}+\dots,\\
		\theta  \rightarrow & h\left(u,\theta\right)-\frac{Pf'}{r}-\frac{1}{2}\ell^{2}\left(h''+\cot\theta h'\right.\\
		& \left.-\frac{1}{2}\csc^{2}\theta\sin2h\right)\frac{1}{r^{2}}+\dots,\\
		\phi \rightarrow &\phi,
	\end{align*}
	with
	\begin{widetext}
		\begin{align*}
			U_{2} & =\frac{\ell^{4}\left(\csc\theta\sin h-h'\right)}{4f'P^{3}}\left[P\left(-2h''-2\cot\theta h'+\csc^{2}\theta\sin(2h)\right)\right.\\
			& \left.-2\left(\csc\theta\sin h-h'\right)\left(f'\left(\partial_{f}P\right)+\left(\partial_{h}P\right)\left(h'+\csc\theta\sin h\right)\right)\right],\\
			R_{0} & =\frac{1}{2}\ell\sin\theta\csc h\left(\frac{\ell\left(-h''-\cot\theta h'+\frac{1}{2}\csc^{2}\theta\sin2h\right)}{f'}-\frac{f'P\left(\partial_{h}P\right)}{\ell}+\frac{2\ell\left(\partial_{f}P\right)\left(h'-\csc\theta\sin h\right)}{P}\right),\\
			R_{1} & =\frac{\ell^{4}\csc^{3}\theta\sin^{3}h}{16f'^{2}P^{5}}\left\{ -2P^{2}\left(\partial_{h}P\right)\left(h'\sin\theta\csc h-1\right)^{2}\left(h'\sin\theta\csc h+1\right)\left(8f'\sin\theta\csc h\left(\partial_{f}P\right)\right.\right.\\
			& \left.+3\left(\partial_{h}P\right)\left(h'\sin\theta\csc h+1\right)\right)+4P^{3}\left(\sin^{2}\theta h'^{2}\csc^{2}h-1\right)\left[4\sin\theta f'\csc h\left(\partial_{f}\partial_{h}P\right)\left(h'\sin\theta\csc h-1\right)\right.\\
			& \left.+2\partial_{h}^{2}P\left(h'^{2}\sin^{2}\theta\csc^{2}h-1\right)+\left(\partial_{h}P\right)\left(\sin\theta\csc^{2}h\left(\sin\theta h''+\cos\theta h'\right)-\cot h\right)\right]\\
			& -8\ell^{2}P\left(\partial_{f}^{2}P\right)\left(h'\sin\theta\csc h-1\right)^{3}\left(h'\sin\theta\csc h+1\right)+8\ell^{2}\left(\partial_{f}P\right)^{2}\left(h'\sin\theta\csc h-1\right)^{3}\left(h'\sin\theta\csc h+1\right)\\
			& +\csc^{2}hP^{4}\left[2h''^{2}\sin^{4}\theta\csc^{2}h+4h''\sin^{2}\theta\left(h'\sin\theta\csc^{2}h\left(\cos\theta-2h'\sin\theta\cot h\right)+\cot h\right)\right.\\
			& +h'\left(h'\sin^{2}\theta\csc^{2}h\left(8h'\sin\theta\left(h'\sin\theta\csc^{2}h-\cos\theta\cot h\right)+4\cos(2h)+\cos(2\theta)-11\right)+2\sin(2\theta)\cot h\right)\\
			& \left.\left.-3\cos(2h)+5\right]\right\} 
		\end{align*}
	\end{widetext}
	and $P=P\left(f\left(u,\theta\right),h\left(u,\theta\right)\right)$.
	The functions $f$ and $h$ are required to satisfy the following conditions
	\begin{equation}
		h'^{2}+\frac{P^{2}f'^{2}}{\ell^{2}}=\csc^{2}\theta\sin^{2}h,\label{eq:condition1}
	\end{equation}
	\begin{equation}
		P\left(\dot{f}h'-f'\dot{h}\right)=\csc^{2}\theta\sin^{2}h.\label{eq:condition2}
	\end{equation}
	Note that in the linear approximation, Eq~(\ref{eq:condition1})
	indicates that $h=\theta$, while Eq.~(\ref{eq:condition2}) implies Eq.~\eqref{eq:condlin}, i.e., $\dot{f}=-\omega$.
	In the flat limit ($\ell\rightarrow\infty$), Eq.~\eqref{eq:condition1} implies that $h=\theta$, while according to Eq.~\eqref{eq:condition2} one has $\dot{f}=P^{-1}$. This condition is the same as that obtained in  \cite{vonderGonna:1997sh}.
	
	\bibliographystyle{ieeetr}
	\addcontentsline{toc}{section}{\refname}\bibliography{asymp-desitter}
	
\end{document}